\documentclass{aa}  

\usepackage{graphicx}
\usepackage{txfonts}
\usepackage{amssymb}
\usepackage{multirow}
\usepackage{float}
\usepackage{gensymb}
\usepackage{xcolor}
\usepackage{verbatim} 
\usepackage{rotating}
\usepackage{booktabs}
\usepackage[]{hyperref}
\hypersetup{backref=true, pagebackref=true, hyperindex=true, breaklinks=true,colorlinks=true,urlcolor=blue, linkcolor=blue, citecolor=blue,pagecolor=red, bookmarks=true, bookmarksopen=true}

\begin{document}

   \title{Planet-forming disks and their environment \\ across regions and time from the full NIR census}

   \author{A. Garufi\inst{1,2}
          \and C. Ginski\inst{3}
          \and M. Benisty\inst{2}
          \and M. Vioque\inst{4}
          \and A. Winter\inst{5}
          \and J. Huang\inst{6}
          \and C.F. Manara\inst{4}
          \and C. Dominik\inst{7}
          }

   \institute{INAF - Istituto di Radioastronomia, Via Gobetti 101, I-40129, Bologna, Italy. 
              \email{antonio.garufi@inaf.it}
        \and Max Planck Institute for Astronomy, Konigstuhl 17, 69117
Heidelberg, Germany
        \and Centre for Astronomy, Dept. of Physics, National University of
Ireland Galway, University Road, Galway H91 TK33, Ireland
        \and European Southern Observatory, Karl-Schwarzschild-Str. 2, 85748 Garching bei München, Germany
        \and Astronomy Unit, School of Physics and Astronomy, Queen Mary University of London, London E1 4NS, UK
        \and Department of Astronomy, Columbia University, 538 W. 120th
Street, Pupin Hall, New York, NY, USA
        \and Anton Pannekoek Institute for Astronomy, University of Amsterdam,
Amsterdam, The Netherlands
             }

   \date{Received -; accepted -}

 
  \abstract
   {The evolution of planet-forming disks and the processes of planet formation influence each other, and both of them are possibly impacted by the local environment. Extensive high-resolution imagery of disks across space and time is the best tool for determining their evolution. We compiled a comprehensive list of disk-bearing young stars with near-IR high-contrast images available. The sample sums up to 268 sources, including 51 targets with no prior publications, which makes this study the largest of its kind and the most extensive release of IR disk images to date. Our census reveals very diverse disk and ambient morphologies. Disks in Lupus are bright, in Chamaeleon are faint, in Corona Australis and Taurus are frequently surrounded by ambient emission. Disks experience an abrupt increase in IR brightness between \mbox{2 Myr} and 5 Myr. The earliest IR disk cavities around single stars arise after 2--3 Myr explaining why are young disks faint in the near-IR, and determining which disks can live longer. Well-known, high-longevity disks ($>$8 Myr) are always bright. Ambient material is detected in more than 20\% of young sources but the fraction drops with time. We find a clear correspondence for the presence of ambient material with the stellar variability, near-IR excess, and mass accretion rate as well as, in turn, with spirals and shadows in disks. Half of the disks with ambient material show spirals while none of them show rings. We therefore propose that the spirals and the disk warps responsible for shadows are generally induced by late infall from the medium, and that this also affects the stellar accretion. The emerging picture proves the fundamental role of the environment for the disk evolution and planet formation. }


   \maketitle
%

\section{Introduction}

Over the past two decades, the study of exoplanets has advanced rapidly, revealing a clear diversity of orbital configurations and planetary properties \citep[e.g.,][]{Howard2010, Dressing2015, Nielsen2019}. In parallel, our understanding of the planet formation has evolved significantly, driven in particular by theoretical and observational studies of young stellar objects (YSOs). One effective approach to constrain the planet formation is to investigate the evolution of the planet-forming disks themselves. These disks are initially deeply embedded in the natal envelope, and this surrounding material dominates the spectral energy distribution (SED) with the rising continuum characteristic of Class I sources \citep{Lada1987}. As the envelope dissipates, the disk becomes optically visible and evolves through the Class II phase which is dominated by a gas-rich, dusty disk. Finally, disks disperse into a gas-poor hybrid or debris disk with the minor IR excess typical of Class III objects. 

The vast majority of the planet-forming Class I and II objects lie in star-forming regions. Our neighborhood hosts several young clusters that are close enough to enable detailed studies of their three-dimensional structure and kinematics thanks to instruments like \textit{Herschel} and \textit{Gaia} \citep[see e.g.,][]{Lada2003, Andre2010, Zucker2019}. These clouds are clearly not homogeneous. For example, Taurus and Lupus form mostly low-mass stars in loose association and filamentary structures \citep[e.g.,][]{Palmeirim2013, Benedettini2015} while Orion hosts multiple dense clusters and high-mass feedback \citep[e.g.,][]{Kounkel2018, Goicoechea2019}, and Ophiuchus is an embedded cluster with a high surface density and several embedded protostars \citep[e.g.,][]{Andre2007}. The complex formation and evolution history of sub-clusters within these clouds is appreciable from the different surface density and fraction of disk-bearing stars. Chamaeleon I and II, for instance, are actively forming stars while Chamaeleon III is not \citep[e.g.,][]{Belloche2011, Tsitali2015}. Upper Scorpius is the youngest sub-group of the Sco-Cen association but is old enough to represent the endpoint of the planet-forming disk phase \citep[e.g.,][]{Pecaut2016}.

An unprecedented view of the morphology of the individual disks in these nearby star-forming regions has been provided by the latest generation of high-resolution imaging ($\lesssim$0.1$\arcsec$), namely the Atacama Large \mbox{(sub-)Millimeter} Array (ALMA) and the high-contrast optical and near-IR (NIR) imaging aided by modern extreme adaptive optics (AO) from 8-m telescopes, such as the Spectro-Polarimetric High-contrast Exoplanet REsearch \citep[SPHERE,][]{Beuzit2019} at the Very Large Telescope (VLT) and the Gemini Planet Imager \citep[GPI,][]{Macintosh2014}. The high performance of the AO and the developments of the post-processing techniques drove multiple large observational campaigns of the scattered light from planet-forming disks \citep[see the review by][]{Benisty2023}. The largest of these programs were SEEDS \citep[Strategic Explorations of Exoplanets and Disks with Subaru,][]{Hashimoto2012, Takami2013}, the SPHERE Guaranteed Time Observations \citep[GTO,][]{Garufi2016, Ginski2016, Benisty2017}, LIGHTS \citep[Large Imaging with GPI Herbig/TTauri Survey,][]{Laws2020, Rich2022}, DARTTS-S \citep[Disks ARound TTauri Stars with SPHERE,][]{Avenhaus2018, Garufi2020a}, and DESTINYS \citep[Disks Evolution Study Through Imaging of Nearby Young Stars,][]{Ginski2020, Ginski2021, Ginski2024,Garufi2024}.

The increasing number of individual disks with high-resolution images available recently motivated a shift in the approach to the data, from the initial detailed analysis of mostly exceptionally bright disks to the demography of the whole, less biased sample. From an initial view where disk sub-structures (spirals, cavities, rings) were routinely detected in disks \citep[e.g.,][]{Muto2012, Canovas2013, Quanz2013b}, the paradigm redirected to an increasing number of faint or small disks where no sub-structures was detectable, especially in presence of outer companions \citep[e.g.,][]{Garufi2017, Ginski2024}, high UV field \citep{Valegard2024}, or stellar masses larger than \mbox{3 M$_\odot$} \citep{Rich2022}. The most common disk geometry, particularly in young objects, turned out to be a self-shadowed disk with no cavity and no detectable spirals or rings \citep{Garufi2017, Garufi2022b}. Furthermore, ambient material around the planet-forming disks in form of outflows, streamers, or complex cloudlets \citep[e.g.,][]{Laws2020, Ginski2021, Huang2022} became increasingly present in NIR images from large samples, as opposite to the old Herbig AeBe stars that were initially probed. Despite the large number of recent work on relatively large sub-samples, the last effort to embrace the whole available NIR sample in one study is almost one decade old \citep{Garufi2018}, and it only included 58 sources.  

In this work, we conduct the full census of available NIR images of planet-forming to date, thus as to gather a sample that is almost five times larger than that of \citet{Garufi2018}. For the first time, the sample is large enough to enable the study of the sub-samples from individual star-forming regions (mainly Taurus, Ophiuchus, Chamaeleon, Lupus, Upper Sco, and Orion), as well as those from systems with different geometry and age. The manuscript is organized as it follows. We describe the sample and calculate the stellar properties in Sect.\,\ref{sec:sample}, then we scrutinize the NIR images and describe characteristics and trends of disk and environmental material in Sect.\,\ref{sec:analysis}, and finally discuss our findings and conclude in Sects.\,\ref{sec:discussion} and \ref{sec:conclusions}.

\section{Sample selection and properties} \label{sec:sample}

We compiled a comprehensive list of sources with a planet-forming disk for which NIR high-contrast images have been taken, to the best of our knowledge. In this section, we describe the sample design and the target properties to lay the groundwork for the analysis of Sect.\,\ref{sec:analysis}.

\subsection{Sample design} \label{sec:sample_publ}
The sample analyzed in this work is composed of both published and original datasets. It mostly includes Class II sources, although we did not set any sharp boundary with the few \mbox{Class I} sources that could be observed with the currently available AO systems. We did not include some embedded Class I objects\footnote{namely Mon R2 IRS 3, Parsamian 21, Elia 2-21, V346 Nor, WW 33a, Elia 2-29, YLW 15, YLW 16A, WL17, WL20.} observed with VLT/NaCo \citep[see][]{deRegt2024} for which there is no optical and near-IR photometry available to constrain the stellar properties, and for which only the pristine envelope is visible from the high-contrast image. For the same reason, we excluded all targets with edge-on disks\footnote{e.g., 2MASS J04202144+2813491, IRAS 04302+2247, DG Tau B.} that strongly veil their host star such that they can only be observed with the Hubble Space Telescope (HST) and James Webb Space Telescope. The constraints that can be obtained for disks in this geometry are very different from those of the rest of the sample, and these are analyzed in dedicated work \citep[see e.g.,][]{Duchene2024, Villenave2024, Tazaki2025}. 

As for the boundary with Class III sources, we discarded any source with no IR excess (see Sect.\,\ref{sec:sample_properties}) detected up to 20 $\mu$m. This criterion is preferred over the traditional near-to-mid-IR color \citep{Lada1987} since a handful of sources with appreciable near-IR excess and a millimeter flux typical of planet-forming disk would appear Class III and be erroneously removed.  

In total, we identified 217 targets with published high-contrast images as of June 2025. A very large fraction of these (170) have (also) been observed with SPHERE. The others have been observed with GPI (24), NaCo (17), Subaru/HiCiao (5), and HST (1). The sample is enriched by 51 sources with SPHERE polarimetric images and no prior publications that are presented in Sect.\,\ref{sec:analysis_new} {and shown in detail in Appendix \ref{appendix:sample}}. Therefore, the final sample sums up to 268 targets. The complete list of these targets is {also} given in Appendix \ref{appendix:sample}.

\subsection{Derivation of target properties} \label{sec:sample_properties}
We calculated stellar mass and age consistently for the entire sample following a standard approach first described in \citet{Garufi2018}. The entire SED of all targets was collected from VizieR\footnote{\url{http://vizier.u-strasbg.fr/viz-bin/VizieR}}, along with some constraints on the effective temperature $T_{\rm eff}$. When available, the BP/RP spectrum of each source was downloaded from the \textit{Gaia} archive\footnote{\url{https://gea.esac.esa.int/archive/}} along with the parallactic measurement to extract the distance $d$. A first flag is raised for targets with an error on the parallax larger than one tenth of the parallax measurement, indicating a critical measurement of $d$. Then, we adopted a PHOENIX model of the stellar photosphere \citep{Hauschildt1999} with $T_{\rm eff}$ and optical extinction $A_{\rm V}$ evaluated from the \textit{Gaia} spectrum and from the literature. A second flag is raised for targets with $A_{\rm V}$ larger than 4, indicating a large uncertainty in the determination of the stellar properties. Several sets of pre-main-sequence tracks \citep{Siess2000, Bressan2012, Baraffe2015, Choi2016, Feiden2016} were used to extract the most likely stellar mass $M_*$ and age $t$ from $T_{\rm eff}$ and from the stellar luminosity $L_*$ measured from the PHOENIX model scaled to the de-reddened $V$ magnitude.

To evaluate the stellar variability, we retrieved a variability amplitude $\Delta V$ defined as the difference between the 95th and 5th percentile $V$ magnitudes from the stellar lightcurve by \mbox{ASAS-SN}\footnote{\url{https://asas-sn.osu.edu/}}. These lightcurves are available for nearly all sources (259), typically span more than 1000 days, and contained several hundred individual measurements. Also, the accretion rate of some more than a half of the sample was collected from the relevant literature \citep[e.g.,][]{Hillenbrand1992, Hartmann1998, Fedele2010, Fairlamb2015, Alcala2017, Gangi2022, Manara2023, Delfini2025}.

Finally, some unresolved disk properties were derived from the IR-to-mm photometry. The near- (1.2--4.5 $\mu$m), mid- (4.5--22 $\mu$m), and far-IR (22--450 $\mu$m) excesses were measured from the SED as the photometric excess over the PHOENIX model of the stellar photosphere, and normalized to the stellar flux. When available, the flux at 1.3 mm $F_{1.3}$ of each target was scaled to the same distance for the entire sample to obtain a crude relative estimation of the disk mass in dust of the individual sources.

\subsection{Regional census} \label{sec:sample_regions}

A large fraction of the 268 sources studied in this work are part of a star-forming region. The most represented regions are Taurus, Upper Scorpius, and Orion, but a relevant number of sources is also available for Chamaeleon, Lupus, Ophiuchus, and Corona Australis (see Table \ref{table:sub_sample}). The 62 objects that are not part of these seven regions are either isolated sources or members of a much smaller sub-sample from other regions such as Perseus and Serpens. The 9 sources from Lower and Upper Centaurus can still be studied statistically by incorporating them in the large Sco-Cen association. In some diagrams of the manuscript, the 6 sources of $\eta$ Cha are also scrutinized as part of their parental region despite their low number.

\begin{table}[h]
	\centering
	\caption{Sub-samples described in Sects.\,\ref{sec:sample_regions} and \ref{sec:sample_stars}.}
	\label{table:sub_sample}
	\begin{tabular}{lcccc} 
		\hline\hline
		Sub-sample & N & $d$ & $M_*$ ($\sigma (M_*)$) & $t$ ($\sigma (t)$)\\
         & & (pc) & (M$_\odot$) & (Myr) \\
		\hline
		Taurus & 48 & 144 & 0.5 (0.7) & 1.3 (1.8) \\
		Chamaeleon & 27 & 190 & 0.9 (0.7) & 1.3 (1.2) \\
        Ophiuchus & 20 & 137 & 1.1 (1.0) & 1.6 (2.6) \\
        Lupus & 23 & 156 & 0.6 (0.6) & 2.1 (2.2) \\
        Corona Australis & 12 & 155 & 1.3 (0.9) & 3.8 (4.6) \\
        Orion & 38 & 387 & 1.8 (1.2) & 4.5 (4.2) \\
        Upper Scorpius & 40 & 143 & 1.0 (0.8) & 5.8 (5.0) \\
		$\eta$ Chamaleontis & 6 & 102 & 1.1 (0.5) & 6.4 (4.3) \\
        Centaurus & 8 & 112 & 1.4 (0.6) & 10.9 (5.5) \\
        \hline
        Others & 47 & 440 & 3.2 (2.4) & 1.0 (4.5) \\
		\hline
        \hline
        Young star & 153 & 169 & 0.9 (2.0) & 1.0 (0.7) \\
        Mature star & 74 & 157 & 1.4 (0.7) & 4.5 (1.3) \\
        Old star & 41 & 146 & 1.5 (0.4) & 11.9 (3.2) \\
        \hline
        \hline
        Low-mass star & 80 & 152 & 0.5 (0.1) & 1.2 (2.6) \\
        Sun-like star & 79 & 154 & 1.0 (0.2) & 3.9 (4.8) \\
        Intermediate-mass & 109 & 310 & 2.2 (1.7) & 2.8 (4.1) \\
        \hline
        \hline
        Single star (*) & 165 & 196 & 1.6 (2.1) & 1.0 (4.5) \\
        Encircled stars (**) & 12 & 189 & 2.0 (1.6) & 1.1 (6.5) \\
        Close system (*)* & 58 & 167 & 1.5 (1.5) & 1.7 (3.2) \\
        Wide system (*) \ * & 33 & 191 & 1.6 (2.4) & 1.8 (2.5) \\
        \hline
    	\end{tabular}
    \tablefoot{Columns are sub-sample, number of sources, median distance, stellar mass (median and standard deviation), and age (median and standard deviation). Sub-samples are: star-forming regions, age (0--3, 3--8, 8--20 Myr), stellar mass (0.3--0.7, 0.7--1.5, 1.5--8 M$_\odot$), multiplicity (single stars, circum-multiple disks -- namely compact stellar systems encircled by a disk -- and stars surrounded by a circumprimary disk with the closest, outer companion at less and more than 300 au). Regions are sorted by the age of our sub-sample.}
\end{table}

To evaluate the level of completeness of the sample, we used the recent census of YSOs in nearby star-forming regions based on \textit{Gaia} data. We focus on six of the seven well-populated regions, excluding Orion, where the vast number of sources \citep[almost 3000,][]{Grossschedl2019} limits any aspiration to completeness in our sample. We retrieved the total number of members and of \mbox{Class II} from Taurus \citep{Esplin2019, Luhman2023}, Upper Sco \citep{Luhman2020a}, Chamaeleon \citep{Galli2021}, Lupus \citep{Luhman2020b}, Ophiuchus \citep{Esplin2020}, and Corona Australis \citep{Galli2020, Rigliaco2025}. Thus, we calculated the fraction of observed Class II sources in each region, as well as the fraction of observable Class II sources that were observed (see Table \ref{table:region_fraction}). Here we consider a source observable based on the criterion of a stellar $R$ magnitude lower than 13, which is considered the current technical limits for the current generation of AO systems\footnote{To promptly find a consistent value for the entire sample, we use the \textit{Gaia G$\rm _{RP}$} passband which is wider in wavelength than the Johnson $R$. The exact conversion depends on the individual color. However, the difference is typically within 1 mag.}. The fraction of observable sources depends on the distance, on the extinction, and on the age (since the stellar luminosity decreases with time). This explains why this fraction is larger in Taurus and Lupus (see Table \ref{table:region_fraction}) than in Chamaeleon (more distant), Ophiuchus and Corona Australis (extincted), and Upper Sco (older).

\begin{figure*}
        \centering
    \includegraphics[width=1.0\linewidth]{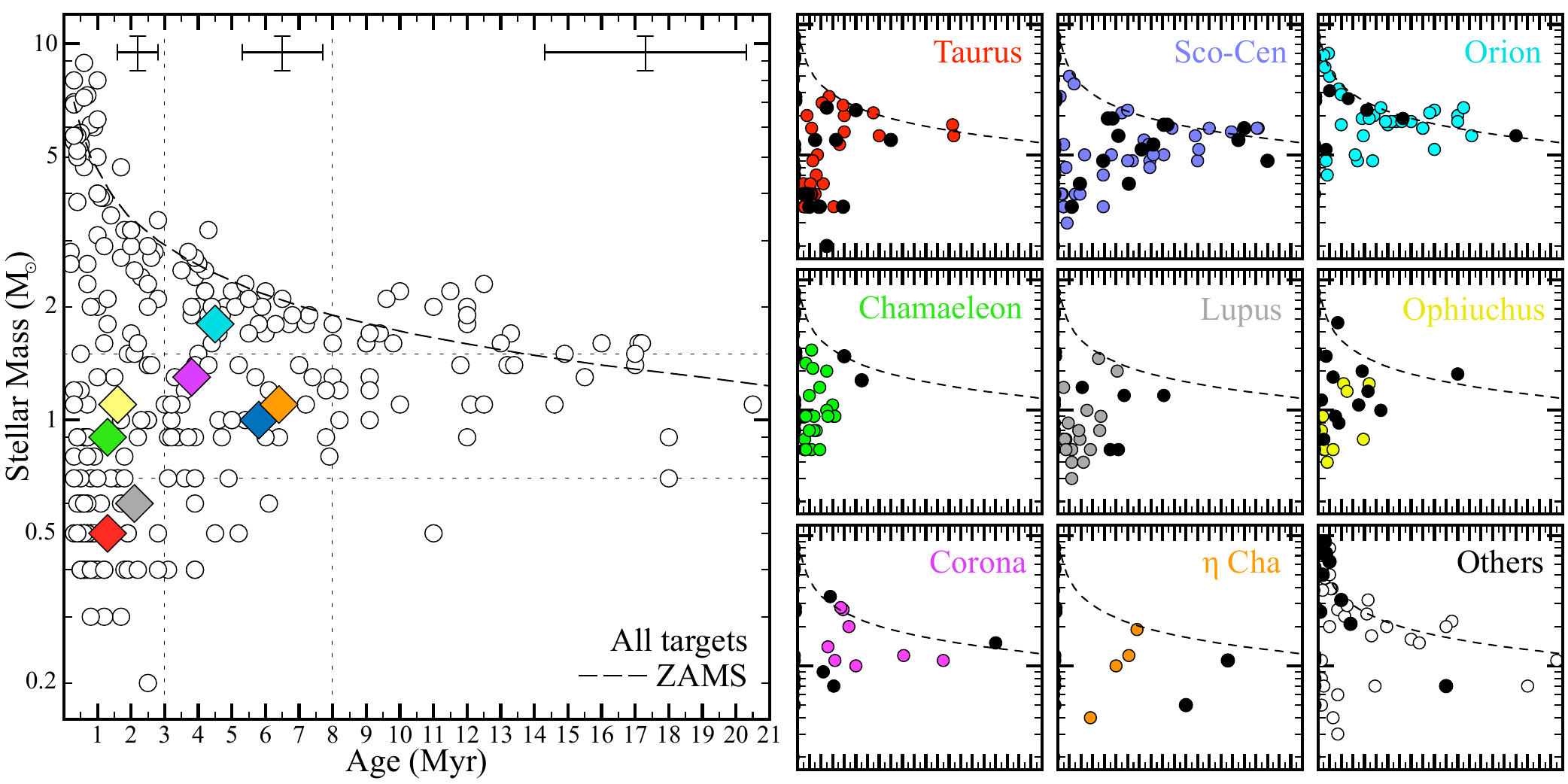}
    \caption{Stellar mass vs age diagram for the entire sample (left) and for the individual star-forming regions (right). The dashed line indicates the zero-age main sequence. In the main diagram, the diamonds indicate the median value of the regions color-coded as in the small panels while the median error bars at the three age stages (see Table \ref{table:sub_sample}) are given to the top. Black points of the small panels indicate flagged sources (see Sect.\,\ref{sec:sample_stars}).}
    \label{fig:mass-age}
\end{figure*}

\begin{table}[h]
	\centering
	\caption{Completeness of the sample in star-forming regions.}
	\label{table:region_fraction}
	\begin{tabular}{lccccc} 
		\hline\hline
	Region & N$_{\rm II}$ & $f_{\rm II/all}$ & $f_{\rm obs/II}$ & $f_{\rm obs/II(R<13)}$ & $f_{\rm II(R<13)/II}$ \\ 	
        \hline
        Taurus & 220 & 45\% &  22\% & $\sim$60\% & $\sim$40\% \\
        Chamaeleon & 99 & 50\% &  27\% & $\sim$90\% & $\sim$30\% \\
        Ophiuchus & 153 & 40\% & 13\% & $\sim$70\% & $\sim$20\% \\
        Lupus & 82 & 55\% &  28\% & $\sim$60\% & $\sim$50\% \\
        Corona Aus & 47 & 20\% & 25\% & $\sim$90\% & $\sim$30\% \\
        Upper Sco & 294 & 20\% &  10\% & $\sim$50\% & $\sim$20\% \\
        \hline
        \end{tabular}
    \tablefoot{Columns are regions, total number of Class II sources (see Sect.\,\ref{sec:sample_regions} for references), fraction of Class II sources in the whole region, fraction of Class II observed, fraction of Class II sources observable with current AO ($R<13$ mag) that have been observed, and fraction of Class II that are observable. {Orion is not included since the high number of sources (few thousands) precludes any aspiration toward completeness.} Compared to \citet{Galli2021}, the N$_{\rm tot}$ of Chamaeleon includes the 15 stars added to the count by \citet{Testi2022}. Regions are sorted by the age of our sub-sample.}
\end{table}

We found that the fraction of Class II sources observed in these nearby regions is still relatively low (from 13\% to 28\%). Nonetheless, the fraction of observable Class II that have been observed spans from 50\% (in Upper Sco) to 90\% (in Chamaeleon and Corona Australis). As discussed by \citet{Garufi2024} for the case of Taurus, these numbers reflect our inability to observe the common low-mass stars (with a typical technical threshold at \mbox{0.4 M$_\odot$}), and are indicative of a rather complete sample for solar-like stars and above, but of a scarce completeness for low-mass stars.

\subsection{Stellar mass and age} \label{sec:sample_stars}
Targets from the whole sample are classified based on the stellar mass and age calculated in Sect.\,\ref{sec:sample_properties}. Adopting some age boundaries based on the disk fraction in stellar clusters \citep[focusing on the comparison with magnetic pre-main-sequence tracks by][]{Richert2018}, we consider stars young up to an age of 3 Myr (that is when the disk fraction decreases to 50\%), mature up to 8 Myr (20\%), and old thereafter. Slightly more than half of the stars in our sample are young, and two-third of the other half are mature (see Table \ref{table:sub_sample}). As for their mass, sources can be divided in three similarly large sub-classes: low-mass (up to 0.7 $\rm M_\odot$), Sun-like, and intermediate-mass stars (from 1.5 $\rm M_\odot$ upward). The overall stellar mass and age distribution is drawn in Fig.\,\ref{fig:mass-age}, offering a good overview of the sample.

In the upper part of the main diagram (intermediate-mass stars), a clear over-density of sources near or at the zero-age main sequence (ZAMS) is visible. Since the temperature and luminosity of stars reaching the ZAMS will not significantly change, these targets may be slightly older than what is determined, although the absence of disk-bearing stars more massive than 2 M$_\odot$ after 3 Myr is also observed in larger sample \citep[e.g.,][]{Ribas2015}. In the lower part of the main diagram (low-mass stars), a scarcity of sources increasing with time is evident. This is due to the decreasing stellar luminosity throughout the evolution of low-mass stars that makes them eventually unobservable by the current generation of AO systems. This technical limit also induces the substantial absence of very-low-mass stars (<0.3 M$_\odot$) that are always too faint. In fact, the only source in the diagram with 0.2 M$_\odot$ mass (ZZ Tau) has been observed from space with HST.  

The age and mass of the sub-samples from the main star-forming regions are shown in the small panels of Fig.\,\ref{fig:mass-age}, and their median is indicated by the colored diamond in the main panel. In the small panels, flagged targets are shown in black. These are sources with large extinction or high \textit{Gaia} parallactic errors (see Sect.\,\ref{sec:sample_properties}) or stars hosting a significantly inclined disk that are reported in Sect.\,\ref{sec:analysis_morphology}. In fact, stars with this disk geometry are known to appear artificially older \citep{Garufi2018, Garufi2022b} likely because the circumstellar extinction exerted by the disk is not properly accounted (unlike the interstellar extinction), and the star appears fainter and thus older. Notably, {a large fraction} of the {fourth-quartile} oldest sources in {Taurus, Chamaeleon, Lupus, Ophiuchus, and $\eta$ Cha} are flagged. In these regions, the quasi totality of the unflagged sources aligns with the regional age {(discussed in Sect.\,\ref{sec:discussion_longevity})}. The only exceptions to this trend are three intermediate-mass stars in Taurus and Lupus (UX Tau, MWC480, and HD142527). In addition, CQ Tau and MWC758 are significantly older than the Taurus population because they belong to an older, peripheral group \citep[see][]{Luhman2023, Garufi2024}. {While flagged sources are retained in the sample, they are sometimes excluded from time-based plots, as specified in the text.}

Instead, both Sco-Cen and Orion display a much larger, genuine spread in age than the younger regions. The overall distribution of sources in the diagram also helps understand why the median stellar mass in these two regions is higher than in the young counterparts. Low-mass stars at their age have become too faint to be observed. This effect is particularly important for Orion which is also lying at a further distance. All in all, Taurus and Lupus are the only regions where we can effectively probe low-mass stars. {The 60\% of their respective sub-sample is made of low-mass stars, resulting in a median of 0.5 and 0.6 $\rm M_{\odot}$} (see grey and red diamonds in Fig.\,\ref{fig:mass-age} {and Table \ref{table:sub_sample}). These regions are in fact} close, young, and not particularly extincted (see Sect.\,\ref{sec:sample_regions}). 

\begin{figure*}
        \centering
    \includegraphics[width=0.67\linewidth]{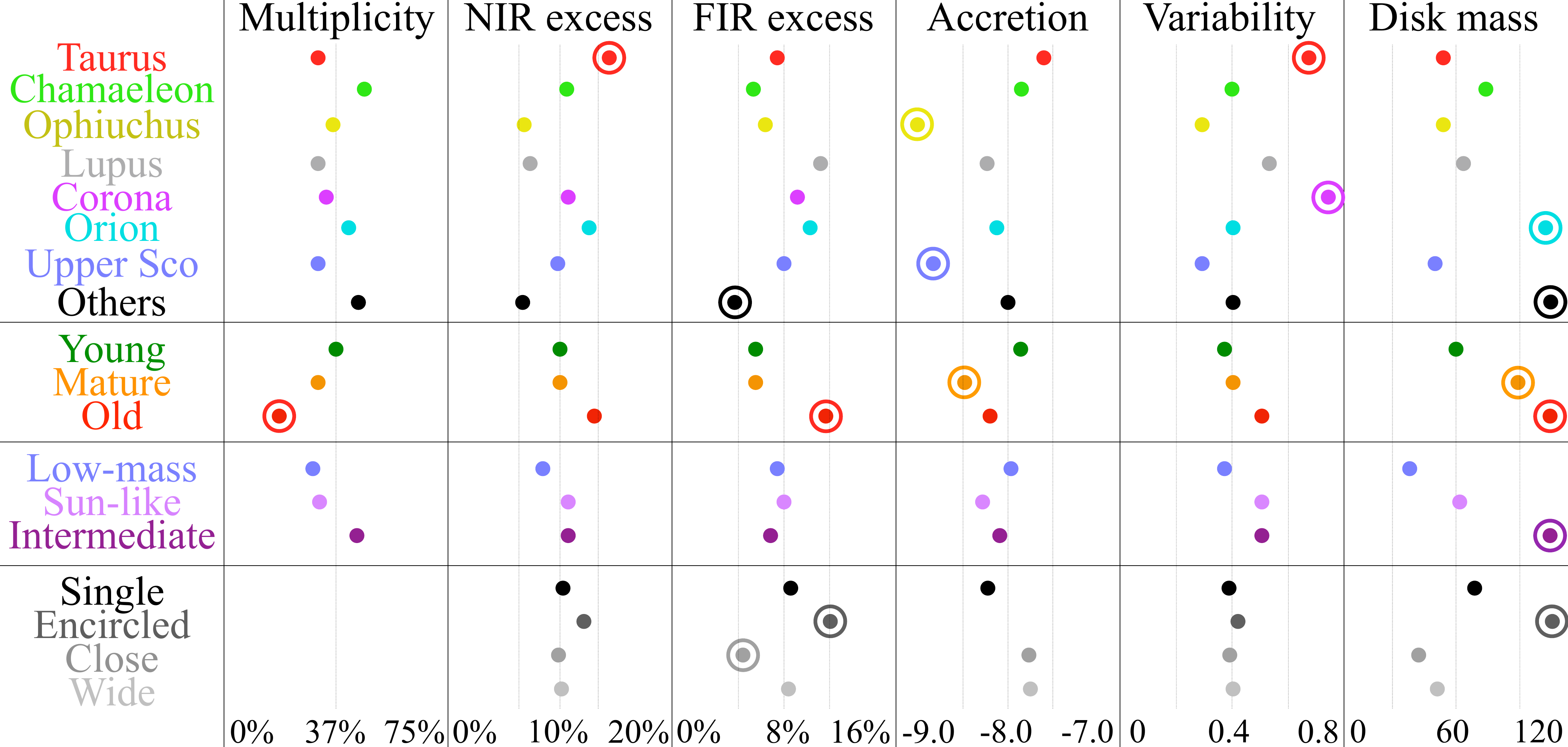}
    \includegraphics[width=0.32\linewidth]{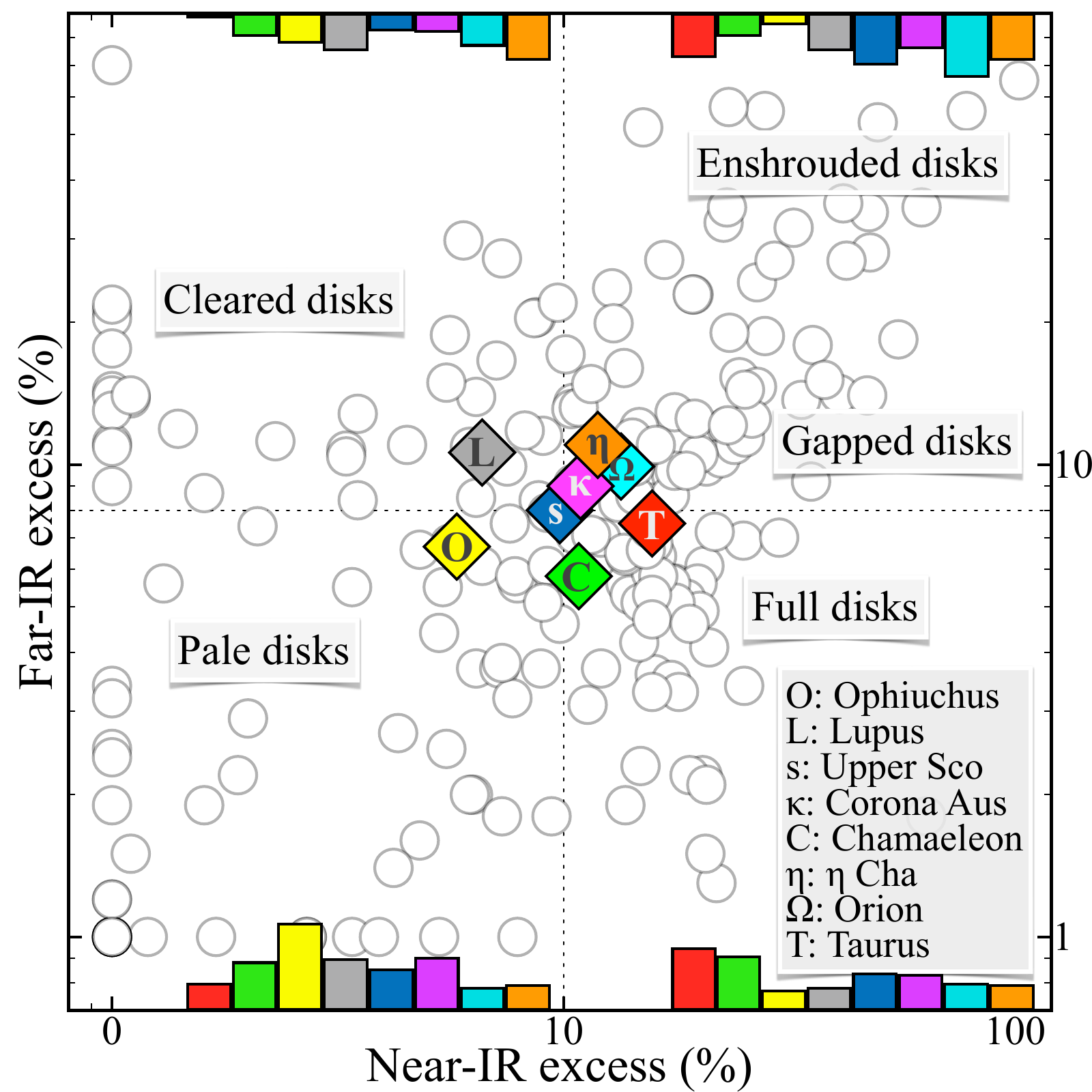}
    \caption{Stellar and disk properties of the sample. Left: median fraction of systems, NIR and FIR excess (fraction of stellar flux), mass accretion rate (log($\rm M_\odot\,yr^{-1}$)), stellar variability (mag), and integrated millimeter flux at a same distance (mJy) for the sub-samples of Table \ref{table:sub_sample}. The median value of the entire sample is at the center of the respective column {while the neighboring vertical lines indicate the 30th and 70th percentiles (for quantities of individual sources only)}. Remarkable values {(smaller and larger percentiles or clear outliers) that are} discussed in the text are highlighted by a circle. Right: FIR vs NIR excess diagram of all sources. The median values of the entire sample determines the four quadrants. The diamonds indicate the median values of Taurus (red), Upper Sco (blue), Orion (cyan), Chamaeleon (green), Lupus (grey), Corona Australis (violet), and $\eta$ Cha (orange). {The vertical bars illustrate the relative number of sources from all regions in each quadrant.}}
    \label{fig:unres_prop}
\end{figure*}

\subsection{Stellar multiplicity} \label{sec:sample_multiplicity}

To the best of our knowledge, approximately one third of the targets are multiple systems. This census is performed through extensive literature review and inspection of the high-contrast image {(see Appendix \ref{appendix:sample})}. Overall, the multiplicity fraction increases with the stellar mass being 28\% for low-mass stars, 32\% for solar-like stars, and 46\% for intermediate-mass stars. This trend is drawn in Fig.\,\ref{fig:unres_prop}, along with several others that will be presented throughout the section. The trend is coarsely consistent with the companion frequency known for main-sequence stars \citep[see e.g.,][]{Duchene2013}. 

Young and mature sources (see Table \ref{table:sub_sample}) exhibit the median multiplicity fraction of the sample while an appreciable drop (from 37\% to 13\%) is found in old ($>$8 Myr) sources. Such a low value cannot be explained solely by the absence of the most massive stars in this age range (see Fig.\,\ref{fig:mass-age}), and it may therefore reflect a real physical effect for which disks in stellar systems hardly live longer than $\sim5$ Myr. The overall binary fraction in our surveyed regions is in line with the global median except in Orion (42\%) and Chamaeleon (48\%). While in Orion we probe higher-mass stars that may naturally explain the increased fraction, the large value of Chamaeleon is peculiar since it can only be partly explained by the stellar mass probed (being still lower than other regions with standard multiplicity).

Thirteen targets in the sample present a geometry where the system is encircled by a circum-multiple disk (CMD), and are either spectroscopic binaries or systems with stars in close orbits surrounded by a large circumbinary or circumtertiary disk (see Sect.\,\ref{sec:analysis_morphology}). The remaining systems are stars with a circumprimary disk and a companion on a wider orbit that we divided between close systems (having the closest companion at separations smaller than 300 au) and wide systems (see Table \ref{table:sub_sample}). The distribution of observed binary separations is also coarsely consistent with that of main-sequence stars, showing a logarithmic plateau between 50 and 500 au.

\subsection{Stellar variability and accretion} \label{sec:sample_variability}

The vast majority of sources show some degree of variability (85\% with $\Delta \rm V \gtrsim0.2$ mag). The median $\Delta \rm V$ value of the entire sample is 0.4 mag (with a large $\sigma$ of 0.7 mag), and as much as 20\% show more than 1.0 mag. There is no substantial difference for the median value across stellar properties (see Fig.\,\ref{fig:unres_prop}). However, there are clear variations from region to region having in particular the Taurus sources a median value of 0.7 mag and the Corona Australis sources as much as 1.0 mag (both with $\sigma$=0.6 mag). This trend is also analyzed in Sect.\,\ref{sec:analysis_ambient}.

The mass accretion rate spans significantly within the sample, with a continuous distribution over five orders of magnitude (from $10^{-5}$ to $10^{-10}$ $\rm M_\odot\,yr^{-1}$), plus an extreme case at \mbox{$10^{-4}$ $\rm M_\odot\,yr^{-1}$} from the outbursting stars Z CMa \citep{Sicilia-Aguilar2020}. To break the known dependence on the stellar mass \citep[e.g.,][]{Muzerolle2003, Natta2004}, we analyze the rate divided by the stellar mass $M_*^{1.8}$. Overall, the median normalized accretion rate decreases with time, although it slightly increases after 8 Myr because of the absence of any old low accretors ($<10^{-9}$ $\rm M_\odot\,yr^{-1}$). In turn, this effect can be related to the absence of any old low-mass stars in our sample (see Fig.\,\ref{fig:mass-age}). The median accretion rate is rather constant across regions except in Upper Sco and Ophiuchus where it is significantly lower ($10^{-8.7}$ and $10^{-8.9}$ $\rm M_\odot\,yr^{-1}$, see Fig.\,\ref{fig:unres_prop}).

\subsection{Disk properties} \label{sec:sample_disks}

The median NIR excess (10\% of the stellar flux) and FIR excess (8\%) of the entire sample determines four quadrants in the FIR-NIR diagram shown in Fig.\,\ref{fig:unres_prop}. As discussed by \citet{Garufi2022b}, sources sitting in the top-right quadrant (high-high) are typically prominent and enshrouded disks with appreciable ambient emission. Cavity-bearing disks with an appreciable inner disk at au-scale (named gapped disks in the figure) also sit in this quadrant. Sources in the top-left quadrant (with FIR-NIR being high-low) are cleared disks with a prominent cavity and no inner disk, while those in the bottom-right (low-high) are full disks experiencing self-shadowing from the inner regions, and those in the bottom-left (low-low) are pale disks that may be undergoing homologous depletion.

We found an appreciably different morphology for sources in different regions. This is actually mostly true for the young Lupus (grey in Fig.\,\ref{fig:unres_prop}), Ophiuchus (yellow), Chamaeleon (green) and Taurus regions (red). On the one hand, several Taurus and Chamaeleon sources tend to sit in the full-disk quadrant with the high NIR and low FIR typical of self-shadowed disks \citep{Garufi2022b}. Chamaeleon in particular shows the smallest median FIR excess of the sample (5.8\%). On the other hand, Lupus and Ophiuchus sources have a low median NIR excess (6.6\% and 5.8\%). While Lupus sources are in turn rather distributed between the cleared-disk and pale-disk quadrants, the majority of the Ophiuchus sources sit in the pale-disk quadrant with both low NIR and FIR excess. These results will be recalled in Sect.\,\ref{sec:analysis_brightness} when analyzing the disk images.

As for the disk mass probed by the millimeter flux, the sample reveals a huge diversity (three orders of magnitude) which is visible across both stellar properties and regions. First, higher disk masses are found around higher mass stars \citep[as known from e.g.,][]{Andrews2013, Pascucci2016}. More interestingly, we observed a clear increase with time (from a median \mbox{60 mJy} at 150 pc for young sources to 115 mJy for old targets) that is discussed in Sect.\,\ref{sec:discussion_longevity}. We also note that the mass of circumprimary disks in systems is lower than in single stars (40 mJy vs 70 mJy) while that of circum-system disks is much larger (190 mJy, although with low statistics).

Finally, the median disk mass in regions shows mild variations (from 50 mJy to 75 mJy, except in Orion with 110 mJy). To evaluate the characteristics of our sources in the context of their respective regions, we compared our median millimeter fluxes with that of the full census from \citet{Andrews2013}, \citet{Ansdell2016}, \citet{Barenfeld2016}, \citet{Pascucci2016}, and \citet{Cieza2019}. We found that our disks are on average more massive than the whole sample by a factor 2--6, with Taurus being the least biased (2.2) and Upper Sco the most biased (6.2). This factor coarsely scales with the median stellar mass probed (Table \ref{table:sub_sample} and Fig.\,\ref{fig:mass-age}) suggesting that the bias on the disk mass is largely inherited from the bias on the stellar mass (see Sect.\,\ref{sec:sample_stars}).

\section{Analysis of the NIR images} \label{sec:analysis}

In this section, we present the NIR high-contrast images, analyze their features, and compare the results with those obtained from the sample discussed in the previous section.

\subsection{Data handling} \label{sec:analysis_data}

\subsubsection{Original SPHERE observations} \label{sec:analysis_new}
Along with the 217 targets from the literature, here we introduce the datasets for 51 sources with no prior publication. In turn, 37 of these are from DESTINYS, 2 from the SPHERE GTO, and 12 from open-time programs (see Appendix \ref{appendix:sample}). In particular, nine targets (of which seven with no prior publications) have been observed in the context of a program (0111.C-0369, PI: Vioque) devoted to probe the scarcely populated regime of young, intermediate-mass stars (age<3 Myr, M$_*=2-3\, {\rm M}_\odot$). 

All data are processed using the same method as for all SPHERE/IRDIS \citep{Dohlen2008} polarimetric data in our possession, as explained in several dedicated papers \citep[e.g.,][]{deBoer2020, Ginski2021, Benisty2023}. Shortly, we employed the IRDIS Data reduction for Accurate Polarimetry pipeline \citep[IRDAP,][]{vanHolstein2020} to generate a set of images in Polarimetric Differential Imaging (PDI). The final FITS file produced is in a standard format agreed among the SPHERE, HiCiao, and GPI communities \citep[see also][]{Rich2022} containing the total intensity $I$, the Stokes parameters $Q$ and $U$, the polarized intensity $PI$, and the radial Stokes parameters $Q_\phi$ and $U_\phi$ \citep[see][]{deBoer2016}. In most cases, we make use of the $Q_\phi$ image since this probes the centro-symmetric component of the polarized light expected in the ideal case of single scattering from a disk around a single star. Instead, the $PI$ image is employed in case of evident material illuminated by more than one star \citep[see][]{Weber2023, Garufi2024}. All images are calibrated in flux following the approach outlined by \citet{Ginski2022}. Some more details on the new datasets are given in Appendix \ref{appendix:sample}.

\subsubsection{Data analysis and scrutiny} \label{sec:analysis_scrutiny}

The first obvious feature to examine in all images is the existence of any detectable signal. This is done by visually inspecting the $Q_\phi$ or $PI$ SPHERE images in our possession, or following the results by authors presenting any NaCo, GPI, or HiCiao images. Our scrutiny determines that some near-IR signal is detected in 192 of the 268 targets (72\%), returning 76 targets that are labeled as non-detections.   

The planet-forming disk is not necessarily detected in all 192 images with signal. In fact, only ambient light is detected in a couple dozen cases either because the disk is small or self-shadowed or because the disk signal is concealed or confused by the ambient signal. With some minor level of subjectivity, our visual inspection determines the detection of a disk in 165 cases (62\% of the total census). The disk brightness and morphology are analyzed in Sects.\,\ref{sec:analysis_brightness} and \ref{sec:analysis_morphology}. Instead, some ambient (non-disk) scattered light can be found in 52 sources (20\% of the total census), and is investigated in Sect.\,\ref{sec:analysis_ambient}.   

From all SPHERE datasets where a disk is detected, we calculated its brightness through the polarized-to-stellar light contrast $\alpha_{\rm pol}$, that is defined as the fraction of average polarized light detected along the disk major axis accounting for its dilution with the separation, and normalized by the stellar flux (see \citealt{Garufi2017} and \citealt{Benisty2023} for details). This measurement reflects the capacity of a stellar photon to reach the disk surface and be efficiently scattered off and polarized, depending thus on the outer disk illumination (primarily) and dust properties (secondarily). This number is available for 140 targets.

\begin{figure}
        \centering
    \includegraphics[width=\linewidth]{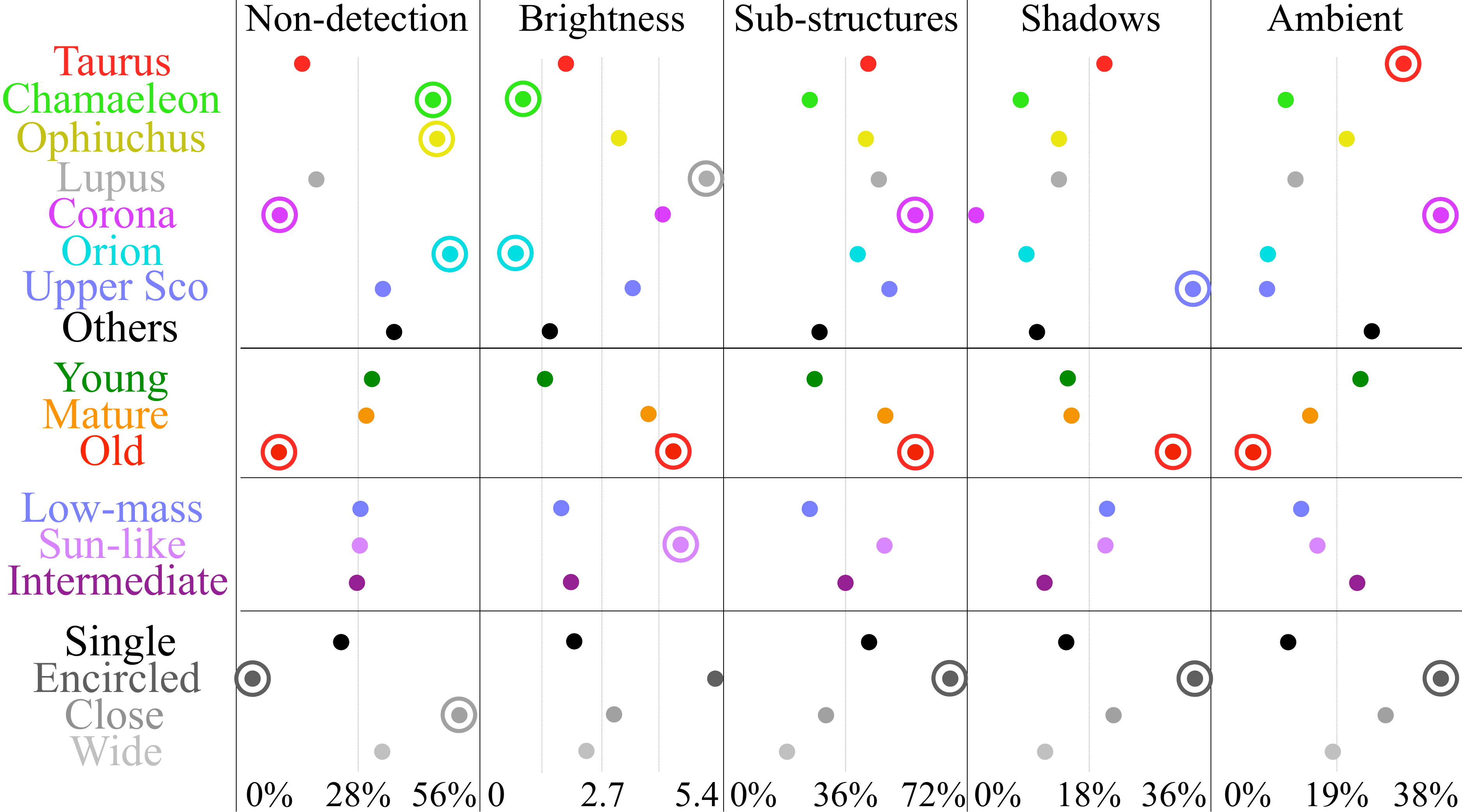}
    \caption{Image properties of the sample. Fraction of non-detections, median disk brightness (in units of $10^{-4}$), fraction of disks with sub-structures {and with shadows} within images with a disk detected, and fraction of images with ambient emission for the sub-sample of Table \ref{table:sub_sample}. Remarkable values discussed in the text are highlighted by a circle.}
    \label{fig:res_prop}
\end{figure}

\begin{figure*}
        \centering
    \includegraphics[width=\linewidth]{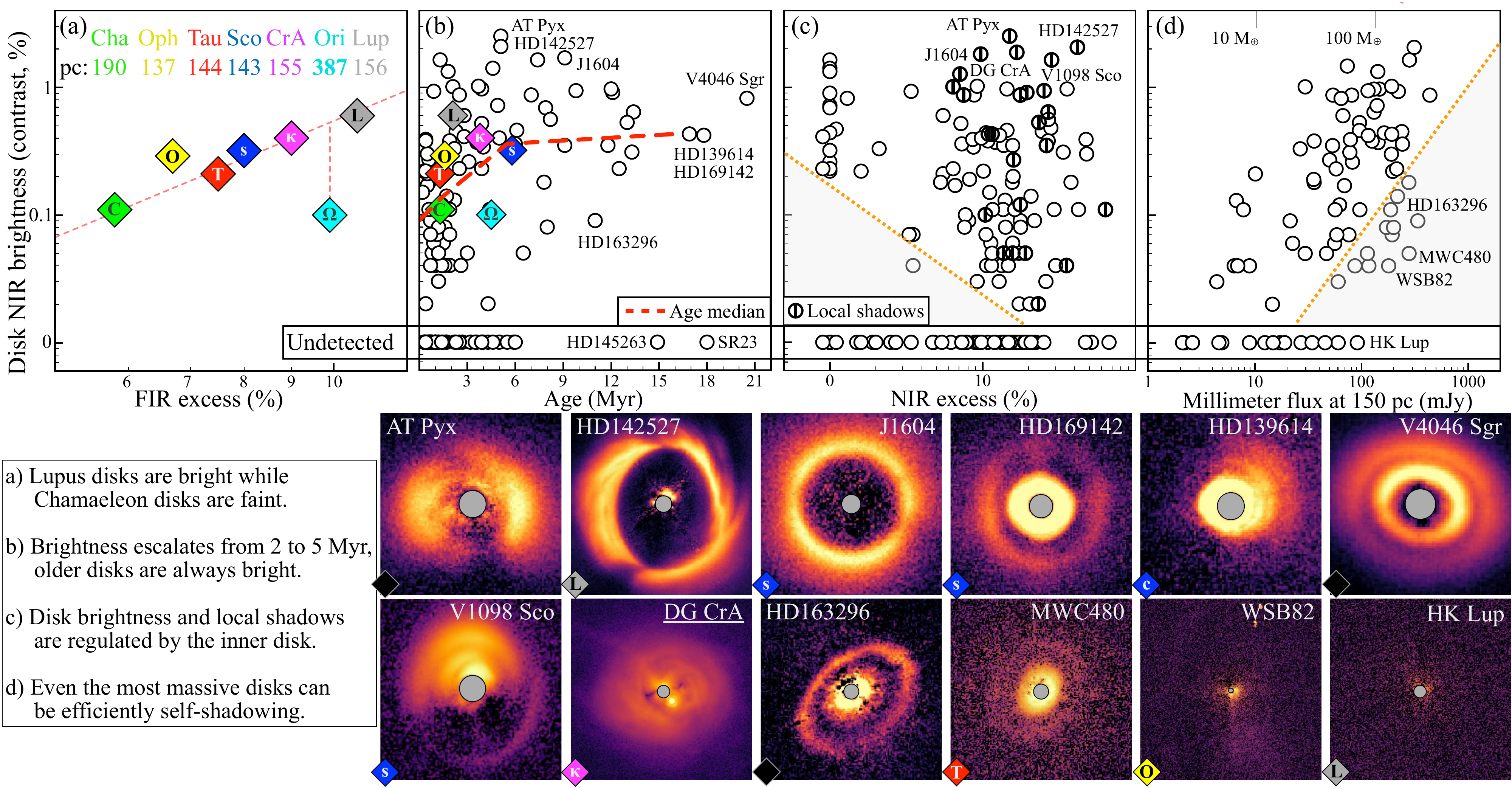}
    \caption{Census of disk brightness. \textit{(a)}: trend with FIR excess for the seven most represented regions. The dashed line is the fit to all points except Orion, which is outlying due to the larger distance. \textit{(b)}: trend with age for all {unflagged}, nearby sources. The median value of all regions (including Orion) is shown with colored diamonds as in (a). The dashed line is the median in different age beams. \textit{(c)}: trend with NIR excess. Sources with local shadows are marked {with crosses. The shaded region indicates the exclusion area, since disks detected in low-NIR sources are always bright.} \textit{(d)}: trend with scaled millimeter flux. The shaded region is what we define the regime of heavily self-shadowed disks. The top x-axis gives a coarse indication of the dust mass corresponding to the measured flux under standard assumptions. The bottom panels are the NIR images of the peculiar targets labeled in the diagrams. {Underlined names indicate images with no prior publications. The colored diamonds to the corner of each image indicate the respective region as in (a), {with the addition of a blue c indicating Centaurus}.}}
    \label{fig:contrast_trend}
\end{figure*}

\subsection{Disk brightness} \label{sec:analysis_brightness}

\subsubsection{Non-detections} \label{sec:analysis_brightness_nondet}
As anticipated in Sect.\,\ref{sec:analysis_scrutiny}, the fraction of non-detections in the sample is 28\%. The vast majority of these sources (78\%) sits in the bottom-left and bottom-right quadrants (pale and full disks) of the diagram of Fig.\,\ref{fig:unres_prop}. This is indicative of the small and self-shadowed nature of these disks being responsible for their non-detections. In fact, with the notable exception of HK Lup \citep{Garufi2020a} and Hn 24 (this work), all the non-detections of the top-right and top-left quadrants (enshrouded and cleared disks) lie at more than 350 pc, suggesting that their distance is the explanation for their missed detection. 

The overall fraction of non-detections (28\%) increases to 54\% in close multiple systems (with companions at less than 300 au, see Sect.\,\ref{sec:sample_multiplicity}) but it drops again to 30\% in wide systems. These numbers most likely reflect the impact of nearby companions on the disk morphology resulting in a large fraction of small disks undetectable in the NIR. Instead, the impact of wider companions is more marginal, and this geometry yields a population of disks comparable to that of single stars.  

Large variations in the fraction of non-detections are also seen across regions (see Fig.\,\ref{fig:res_prop}). In particular, only Upper Sco shows a value in line with the whole sample. Taurus, Lupus, and Corona Australis show smaller values (8\% to 18\%) while Ophiuchus, Chamaeleon, and Orion show larger values (44\% to 49\%). The high value of Orion is explained by the large distance to the region (see below), while that of Ophiuchus is again related to the large fraction of pale disks in the sub-sample (discussed in Sect.\,\ref{sec:discussion_regions}). The high fraction of non-detections in Chamaeleon is less readily explained but it is likely related to the large fraction of multiple systems in the relative sub-sample (see Sect.\,\ref{sec:sample_multiplicity}). {All in all, the detection rate is intimately related with the average disk brightness (see Sect.\,\ref{sec:analysis_brightness_det}) and is further discussed in Sect.\,\ref{sec:discussion_regions}.}

\subsubsection{Detections} \label{sec:analysis_brightness_det}
Considering detections only, the median disk brightness of the whole sample (defined by the $\alpha_{\rm pol}$ of Sect.\,\ref{sec:analysis_scrutiny}) is 0.26\%, which formally separates bright and faint disks. $\alpha_{\rm pol}$ shows a continuous distribution from 0.02\%, imposed by the sensitivity limit for low-mass stars, to 2.50\%. The tendency for inclined disks (more than 60$\degree$) to appear brighter \citep{Garufi2018, Garufi2022b} because of the smaller scattering angles probed and of the partial veiling experienced by the host star is confirmed in this large sample. Their median brightness is in fact almost twice as much as that of less inclined disks, and as many as 8 of the 15 brightest disks are inclined. To remove this bias, we therefore ignore these objects throughout the rest of this section. By doing so, the brightest disks ever imaged in the NIR are, according to our calculations, AT Pyx \citep{Ginski2022}, HD142527 \citep{Avenhaus2017}, and RX J1604.3-2130 \citep[J1604 for simplicity,][]{Pinilla2015b}. The Lupus disks are significantly brighter than the rest of the sample (0.60\%). This is not surprising after acknowledging that their median IR excesses in Fig.\,\ref{fig:unres_prop} sit well within the cleared disk quadrants. On the contrary, the Orion and Chamaeleon disks are, similarly to the non-detection rate, more elusive (0.10\%--0.11\%, see Fig.\,\ref{fig:res_prop}).

In Fig.\,\ref{fig:contrast_trend}, we compare the disk brightness with four other quantities. First, the brightness shows the known, mild correlation with the FIR excess \citep{Garufi2018, Garufi2022b} since both quantities relate with the illumination of the outer disk traced by scattered light and thermal light, respectively. Interestingly, the large scatter in the correlation is significantly reduced by looking at the median value of the available star-forming regions, as is shown in Fig.\,\ref{fig:contrast_trend}(a). The only region that does not follow the trend is Orion which is more than twice more distant. Current-generation NIR telescopes cannot image a larger portion of the disks from Orion whereas the FIR excess is equally measured at that distance. According to the plot, up to 80\% of the scattered light that would be detected at a distance of 150 pc is lost from Orion. For this reason, sources more distant than 300 pc are also removed from the following plots.

The evolution of the disk brightness with time is shown in Fig.\,\ref{fig:contrast_trend}(b). From this diagram, it is evident how the typical disk brightness abruptly increases in a time interval from 1--2 Myr to 3--5 Myr. Young disks show a median brightness of 0.14\% while mature disks of 0.36\% (see also Fig.\,\ref{fig:res_prop}). Nonetheless, the four similarly young regions of our sample (Chamaeleon, Taurus, Ophiuchus, and Lupus) exhibit a very different median value (from 0.11\% to 0.60\%). {Thus, considering the detection rate from Sect.\,\ref{sec:analysis_brightness_nondet}, Lupus disks are typically detected and bright, Taurus disks are detected but faint, and Chamaeleon disks are rarely detected and faint. This tripartition reflects three different disk geometries, namely cleared disks (Lupus), extended self-shadowed disks (Taurus), and more compact disks (Chamaeleon, somehow similar to Ophiuchus disks), and is further discussed in Sect.\,\ref{sec:discussion_regions}.} The other interesting aspect of Fig.\,\ref{fig:contrast_trend}(b) is that disks that are older than 10 Myr are systematically bright, with HD163296 \citep{Monnier2017, MuroArena2018} being a sort of last outpost of the faint disks, and with no undetected Class II disk\footnote{Two non-detections are present in the old regime of Fig.\,\ref{fig:contrast_trend}(b). HD145263 and EM$*$ SR23 are however two quasi-Class III with IR excess only appearing after 5 $\mu$m that were still included in the sample based on the considerations of Sect.\,\ref{sec:sample_publ}.} being older than 6 Myr. This trend also explains the larger median brightness observed in solar-like stars (see Fig.\,\ref{fig:res_prop}), which is partly a bias. In fact, only this type of stars are observed at old ages (see Fig.\ref{fig:mass-age}). Disks around young solar-like stars are still mildly brighter than lower-mass stars but this is also due to the lower contrast that can be measured from more luminous stars.

The interplay between the disk brightness and the inner disk geometry probed by the NIR excess is shown in Fig.\,\ref{fig:contrast_trend}(c). From smaller samples \citep[e.g.,][]{Garufi2022b}, these two quantities anti-correlate because a high NIR excess implies a prominent disk inner edge that casts a shadow on the outer disk. {This} diagram shows a desert to the bottom-left rather than any {strong} correlation. {The desert indicates that disks detected in low-NIR sources are always bright because there is no substantial inner disk that leaves the outer disk in penumbra.} In the figure, we also highlight those sources showing the presence of local shadows (see Sect.\,\ref{sec:analysis_morphology_shadows}) such as an azimuthally confined dark lane \citep[see e.g.,][]{Marino2015, Stolker2016a} or stark brightness variations between two semicircles \citep[e.g.,][]{MuroArena2020, Zurlo2021}. A large fraction of these objects lie to the top-right of the diagram meaning that they are bright in scattered light and exhibit an extremely high NIR excess that must have an extraordinary origin (discussed in Sect.\,\ref{sec:discussion_infall}). Extreme targets in this direction are the aforementioned AT Pyx, HD142527, and J1604, as well as V1098 Sco \citep{Williams2025} and DG CrA (that is reported in this work for the first time). Non-detections are distributed across the entire NIR range. A likely explanation for the non-detection of some low-NIR sources is that their disk is too small to be imaged by current telescopes. 

Finally, the relation with the disk dust mass is explored in Fig.\,\ref{fig:contrast_trend}(d). This diagram shows that more massive disks can be brighter because they are more extended, and any flux detected at large radii, by construction of $\alpha_{\rm pol}$, significantly contributes to the disk brightness. However, it also shows a regime (shaded in the figure) of very massive disks that are very faint in scattered light. These are the prototypical self-shadowed disks discussed by \citet{Garufi2022b}. Extreme cases are the aforementioned HD163296 as well as MWC480 \citep{Kusakabe2012} and WSB82 \citep{Garufi2022b}, although even more intriguing are the non-detection of the massive disk of HK Lup \citep{Garufi2020a} and V1787 Ori \citep{Valegard2024}. Non-detections are found in a large range of millimeter fluxes, reflecting the two aforementioned explanations for a non-detection of \mbox{Class II} disks in scattered light, that are being smaller than $\sim$20 au (at 150 pc) or being efficiently self-shadowed.

\subsection{Disk morphology} \label{sec:analysis_morphology}

\subsubsection{Disk inclination} \label{sec:analysis_morphology_inclination}
Obviously, the first attribute to determine the disk appearance in the image is the disk inclination $i$. While the precise measurement of $i$ solely bases on near-IR imaging can be challenging, a coarse assessment can be done with some basic geometric evaluations in practically all detections. In this work, we lend importance to inclined disks ($i\gtrsim60\degree$) since this geometry biases both the stellar age and disk brightness (see Sects.\,\ref{sec:sample_stars} and \ref{sec:analysis_brightness}). We labeled detected disks as inclined in 28\% of the cases (32\% among the bright disks). These numbers are in line with the expectation (32\%) for disks between 60\degree\ and 75\degree\ from a random distribution of $i$ up to 75\degree. In fact, only a few very inclined disks ($i\gtrsim75\degree$) are part of our sample (see Sect.\,\ref{sec:sample_publ}).

\begin{figure*}
        \centering
    \includegraphics[width=\linewidth]{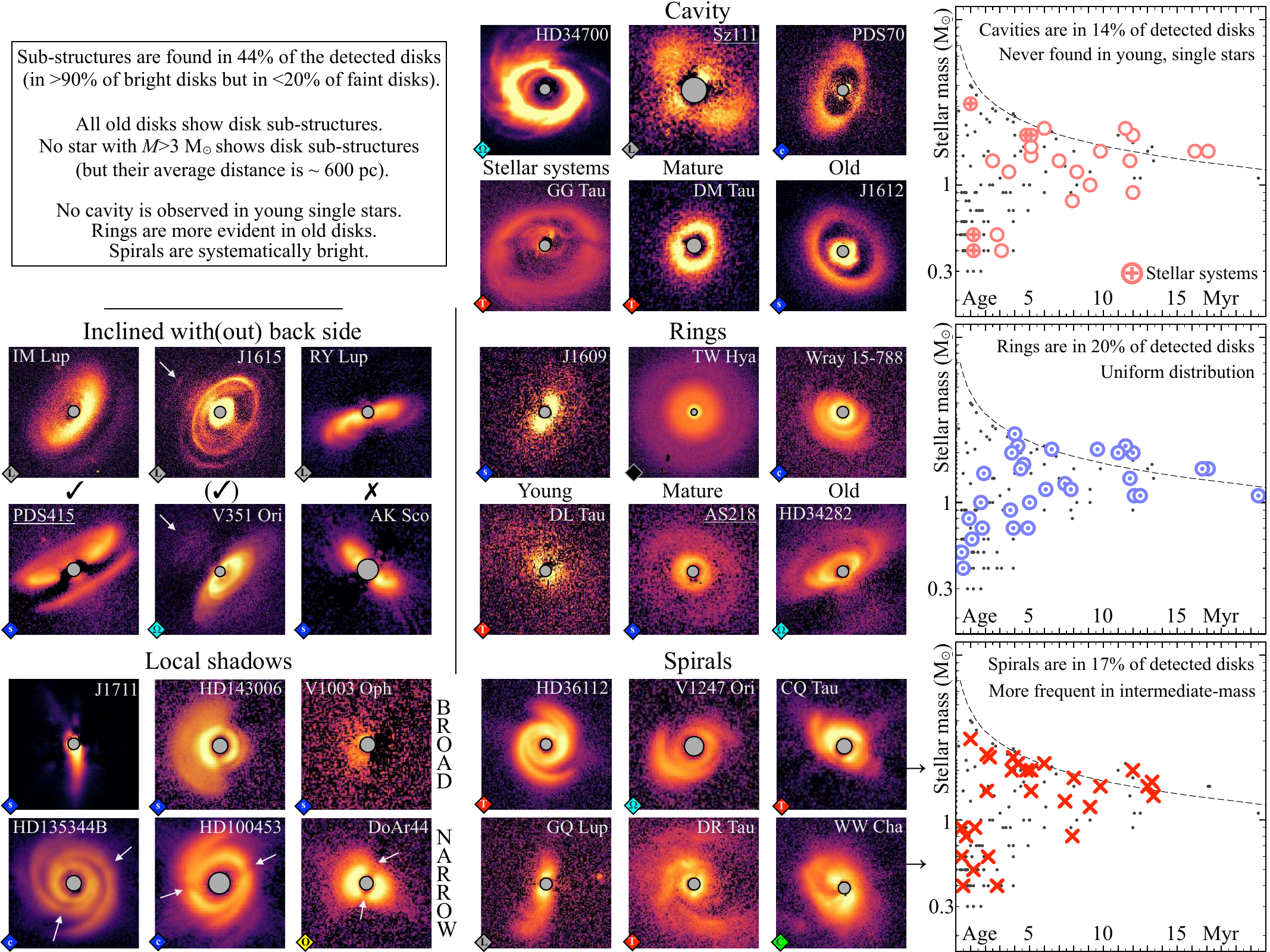}
    \caption{Census of disk sub-structures. The analysis of Sect.\,\ref{sec:analysis_morphology} is visually summarized. The stellar mass vs age diagrams for sources with disk cavity, rings, and spirals are shown to the right (see Fig.\,\ref{fig:mass-age} for details). {Underlined names indicate images with no prior publications. The colored diamonds to the corner of each image indicate the respective region as in Fig.\,\ref{fig:contrast_trend}}.}
    \label{fig:substructures}
\end{figure*}

In some cases with favorable inclination, the disk back side is visible (see some examples in Fig.\,\ref{fig:substructures}). This detection can be useful to constrain the dust scattering and disk geometry through modeling or to measure directly the scale height of the brightest disks. Based on our visual inspection, this is visible in approximately a half of the inclined disks (plus in some less inclined disks such as in IM Lup and RX J1615, \citealt{Avenhaus2018}). The non-detection of a disk back side may in principle be attributed to an insufficient inclination or brightness although some surprising non-detections from bright and very inclined disks exist (e.g., RY Lup and AK Sco, see \citealt{Langlois2018} and \citealt{Garufi2022b}). Radiative transfer simulations \citep{George2025} show that several factors, such as dust mass, grain distribution, and disk stratification, influence the observability of the disk back side.

\subsubsection{Shadows} \label{sec:analysis_morphology_shadows}
As discussed in Sect.\,\ref{sec:analysis_brightness} and shown in Fig.\,\ref{fig:contrast_trend}, some disks exhibit evidence of local shadows. Based on our scrutiny, these are visible in {29} sources ({18\%} of the detected disks). This fraction seems to increase in old sources ({28\%}) compared to young and mature sources ({15\%}), although this may partly be due to the increased brightness of older disks that favors their detection. In particular, we observe a very high fraction of shadows in the old Sco-Cen sources (as much as 35\% of the detections, whereas {only Taurus slightly exceeds the global median value of 17\%, see Fig.\,\ref{fig:res_prop})}.

The only other quantity that correlates with the presence of shadows is the stellar variability (see Sect.\,\ref{sec:sample_variability}). In fact, the median for stars with and without shadows in the disk is 0.7 and 0.4 mag (average 0.90 vs 0.65 mag). This connection is not surprising as the presence of shadows is due to a misalignment or warp in the disk inner regions that is turn related to some perturbing processes also possibly inducing high stellar variability. Further trends pertinent to this results are shown in Sect.\,\ref{sec:analysis_ambient} and discussed in Sect.\,\ref{sec:discussion_infall}. 

The degree of misalignment between inner and outer disk regions determines the azimuthal extent of the shadow \citep[see e.g.,][]{Benisty2017, Nealon2019}. The known dichotomy between narrow ($\lesssim10\degree$) and broad ($\gtrsim90\degree$) shadows is clearly visible from our census (see some examples in Fig.\,\ref{fig:substructures}). Interestingly, these two sub-categories carry the same number of sources. No obvious difference in the stellar and disk properties is found between the two categories.

\subsubsection{Sub-structures} \label{sec:analysis_morphology_substructures}
Sub-structures are arguably the major point of interest in the study of planet-forming disks. An in-depth analysis of their architecture is deferred to a forthcoming publication. In this work, we only examine their frequency and trends with other properties. The first number to be constrained is their overall incidence within detected disks. From our census, this is 44\%. This number is lower than what was constrained from the first, more biased surveys \citep[e.g.,][]{Garufi2018}, and reflects the high number of marginal disk detections in our sample. In fact, 72\% of the bright disks show sub-structures, and two-third of the remaining 28\% are inclined disks with unfavorable configurations to detect them \citep{Dong2016}. Among faint disks, the sub-structure rate drops to 19\%. The obvious connection between detectable sub-structures rate and disk brightness is also clear from Fig.\,\ref{fig:res_prop}, where these two quantities vary similarly across regions, stellar age, mass, and multiplicity. We acknowledge in particular the high sub-structure fraction in Corona Australis (57\%), in old sources (58\%), and in circum-system disks (69\%). On the contrary, there is almost no star more massive than 3 M$_\odot$ where disk sub-structures are detected (with the only candidate being HD58647 imaged with NaCo by \citealt{Cugno2023}). This shortage is however partly due to the large distance of these stars (with a median $d$ of 600 pc), and in particular for stars more massive than 5 M$_\odot$ (that are all further away than 350 pc).

Broadly speaking, disk sub-structures visible from high-contrast imaging can be an inner cavity, concentric rings, and spiral arms (see the census of Fig.\,\ref{fig:substructures}). From our survey, an inner cavity is clearly detected in the NIR from 23 targets (14\% of the detected disks). Cavities smaller than 20 au may be present but concealed by the coronagraph that is typically employed in high-contrast imaging. Overall, the disks with cavities detected are old (as their median age is 7 Myr vs 2 Myr from the complementary sample), massive (their normalized mm flux is 131 mJy vs 56 mJy), and bright in scattered light (with a contrast of 0.87\% vs 0.21\%). Their old age is also appreciable from the notion that as many as 10 of the 23 disks showing a cavity are from the old Sco-Cen association, and none from the young Chamaeleon. 

A sub-class of cavity-hosting disks is that of the giant, circum-multiple disks of HD34700 \citep{Monnier2019}, \mbox{GG Tau} \citep{Itoh2014}, 2MASS J17110392-2722551 \citep{Zurlo2021}, HD142527 \citep{Avenhaus2017}, and GW Ori \citep{Kraus2020}, where most likely their morphology and evolution is primarily driven by their interaction with (and between) the stellar companions. These objects are younger than the whole class of cavity-bearing disks (average 2.6 Myr). Ignoring these, our sample shows no disk with a cavity younger than 2.5--3 Myr (DoAr 44 and Sz111 in Ophiuchus and Lupus, see Fig.\,\ref{fig:substructures}). Therefore, the increasing fraction of cavity-bearing disks with time cannot be solely explained by survival considerations (only disks with prominent sub-structures live long). Instead, it is also due to a real inefficiency for disks to have their cavity sufficiently sculpted in small dust during their first \mbox{3 Myr}, unless a stellar companion is present.  

\begin{figure*}
        \centering
    \includegraphics[width=\linewidth]{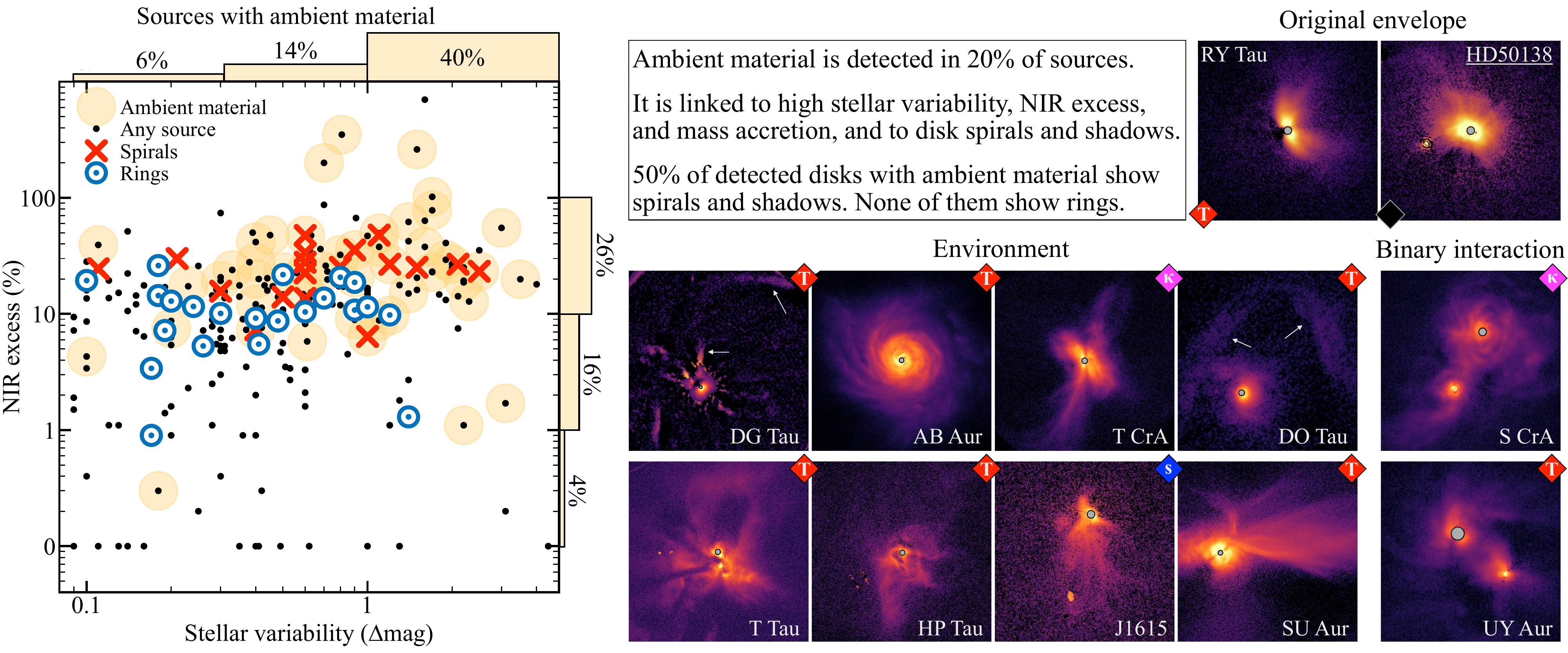}
    \caption{Census of ambient material. The analysis of Sect.\,\ref{sec:analysis_ambient} is visually summarized. The stellar variability vs NIR excess diagram is shown to the left. Faint, ambient structures visible from the images are indicated by arrows. The outermost image regions of DG Tau is in total intensity, and it illustrates the possible origin of the streamer reported by \citet{Garufi2022a}. {Underlined names indicate images with no prior publications. The colored diamonds to the corner of each image indicate the respective region as in Fig.\,\ref{fig:contrast_trend}}.}
    \label{fig:ambient}
\end{figure*}

Rings are detected in 32 disks (20\% of the detections). Spirals are detected in 28 disks (17\%). Rings appear in a rather uniform distribution of stellar mass and age (with a possible under-abundance around young intermediate-mass stars). As shown in Fig.\,\ref{fig:substructures} for a few examples, rings are increasingly evident with time as the disk brightness increases (see Sect.\,\ref{sec:analysis_brightness}). Unlike rings, detected spirals are routinely bright in scattered light. Spirals were thought to be mainly associated with old intermediate-mass, luminous stars \citep{Garufi2018, vanderMarel2021}. In this sample, we however find several spiral-disks around low-mass, young stars (nine below 1 M$_\odot$, see Fig.\,\ref{fig:substructures}). In many of these, the spiral excitation is possibly related to the interaction with companions (GG Tau, GQ Lup) or with the environment (DR Tau, WW Cha, see Sect.\,\ref{sec:analysis_ambient}). In any case, spirals are still preferentially found around intermediate-mass stars, since the fraction of low-mass stars with disk spirals detected is 6\% while this fraction is as much as 20\% in the mass range between 1.5 and 3 M$_\odot$.

Objects with disk spirals and rings do not show any significant difference in disk mass (131 vs 142 mJy), FIR excess (15\% vs 13\%), nor disk brightness (0.46\% vs 0.38\%). However, they do show a clear difference in both the NIR excess from the SED (25\% vs 10\%) and stellar variability (0.6 vs 0.3 mag). Since these two quantities were also found to be high in presence of disk shadows (see Sects.\,\ref{sec:analysis_brightness} and \ref{sec:analysis_morphology_shadows}), an association between shadows and spirals is plausible. In fact, 10 of 28 disks with spirals (36\%) show shadows, while these are only seen in 6 of 32 disks with rings (18\%). These connections will be further investigated in Sect.\,\ref{sec:analysis_ambient}.

\subsection{Ambient morphology} \label{sec:analysis_ambient}

As mentioned in Sect.\,\ref{sec:analysis_scrutiny}, 52 sources (20\% of the total census) show some ambient signal from their image. Two-third of these sources sit in the top-right quadrant (that of enshrouded disks) in the NIR-FIR diagram of Fig.\,\ref{fig:unres_prop}. There is a clearly decreasing trend for the fraction of sources with ambient material with time, since 24\% of young sources, 16\% of mature sources, and 2\% of old sources (which is barely one, J1615-1921 in Upper Sco) show ambient emission (see Fig.\,\ref{fig:res_prop}). However, the stellar region showing the largest incidence of ambient light is Corona Australis (as much as 58\%, see Sect.\,\ref{sec:analysis_maps}), that is not particularly young ($\sim$4 Myr old for our sub-sample, see Sect.\,\ref{sec:discussion_longevity}). Also Taurus shows a significantly higher value than the rest of the sample (28\%), while all others spans from 8\% (Upper Scorpius and Orion) to 20\% (Ophiuchus). Any trend with the stellar mass is much weaker as intermediate-mass stars only show a marginally larger value (23\%) than low-mass stars (15\%). Finally, stellar systems evidently show larger value than single stars (see Fig.\,\ref{fig:res_prop}) because of the interaction between binaries (discussed below).

The ambient signal is seen in a plethora of different {brightnesses and morphologies}. Some examples are shown in Fig.\,\ref{fig:ambient}. We classify sources with ambient material in three main categories that are not mutually exclusive. A few sources (8 from our count) show what appears as leftover material from the envelope or the outflow shell (such as HD45677, DO Tau, and RY Tau by \citealt{Laws2020}, \citealt{Huang2022}, and \citealt{Garufi2019}). A larger number of sources (25) exhibit material that seems connected to both the central star and a stellar companion, indicating a likely binary interaction. Notable examples are \mbox{UX Tau}, \mbox{FU Ori}, and \mbox{S CrA} \citep{Menard2020, Weber2023, Zhang2023b}.  Finally, the largest portion of sources (44) reveals structures such as streamers (e.g., DG Tau, T Cra, and SU Aur by \citealt{Garufi2024}, \citealt{Rigliaco2023}, \citealt{Ginski2021}) or more complex patterns (J1615-1921, T Tau, and HP Tau by \citealt{Garufi2020a}, \citealt{Garufi2024}) indicating a generic interaction with the environment.   

\begin{figure*}
        \centering
    \includegraphics[width=13cm]{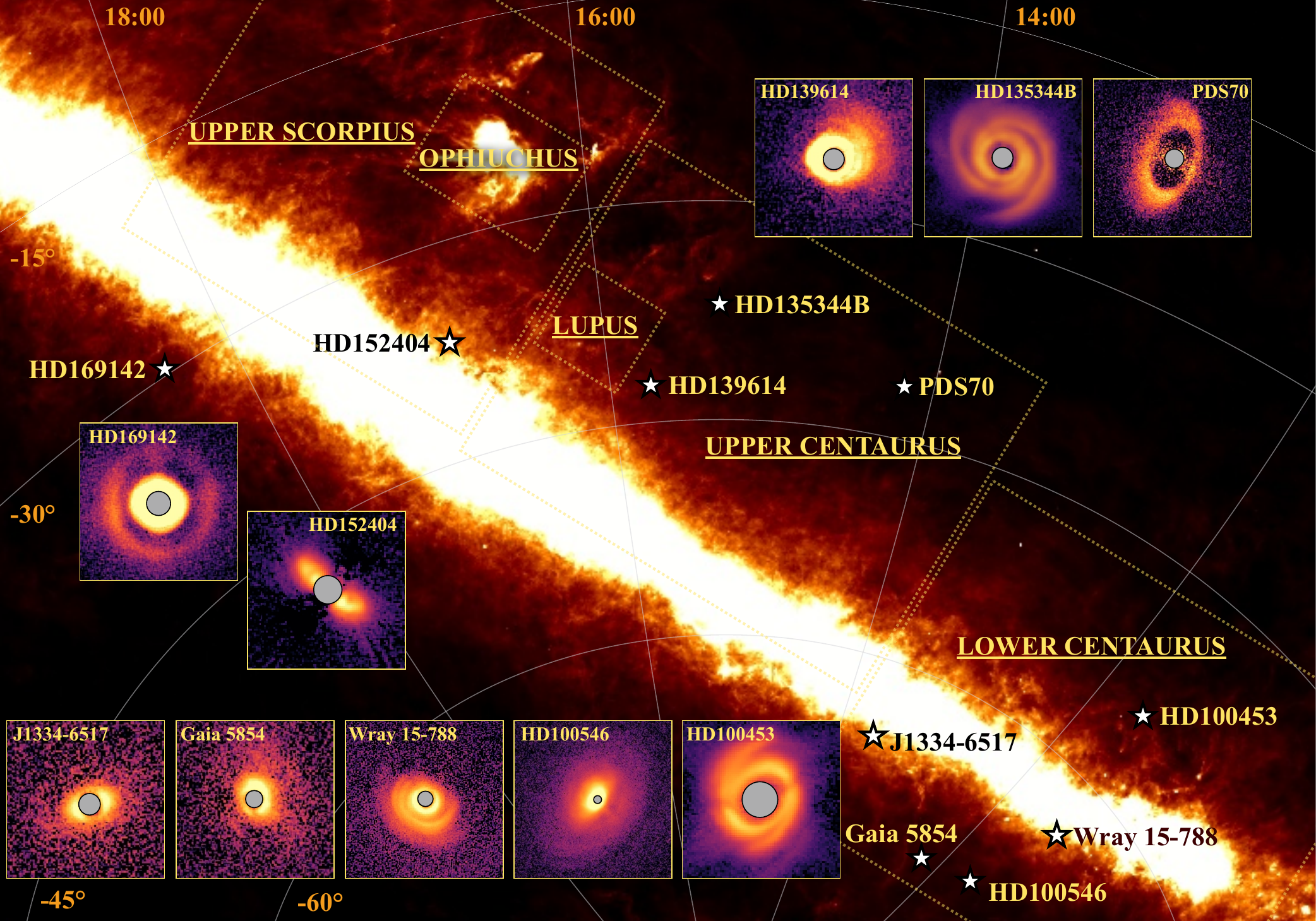}
    \includegraphics[width=13cm]{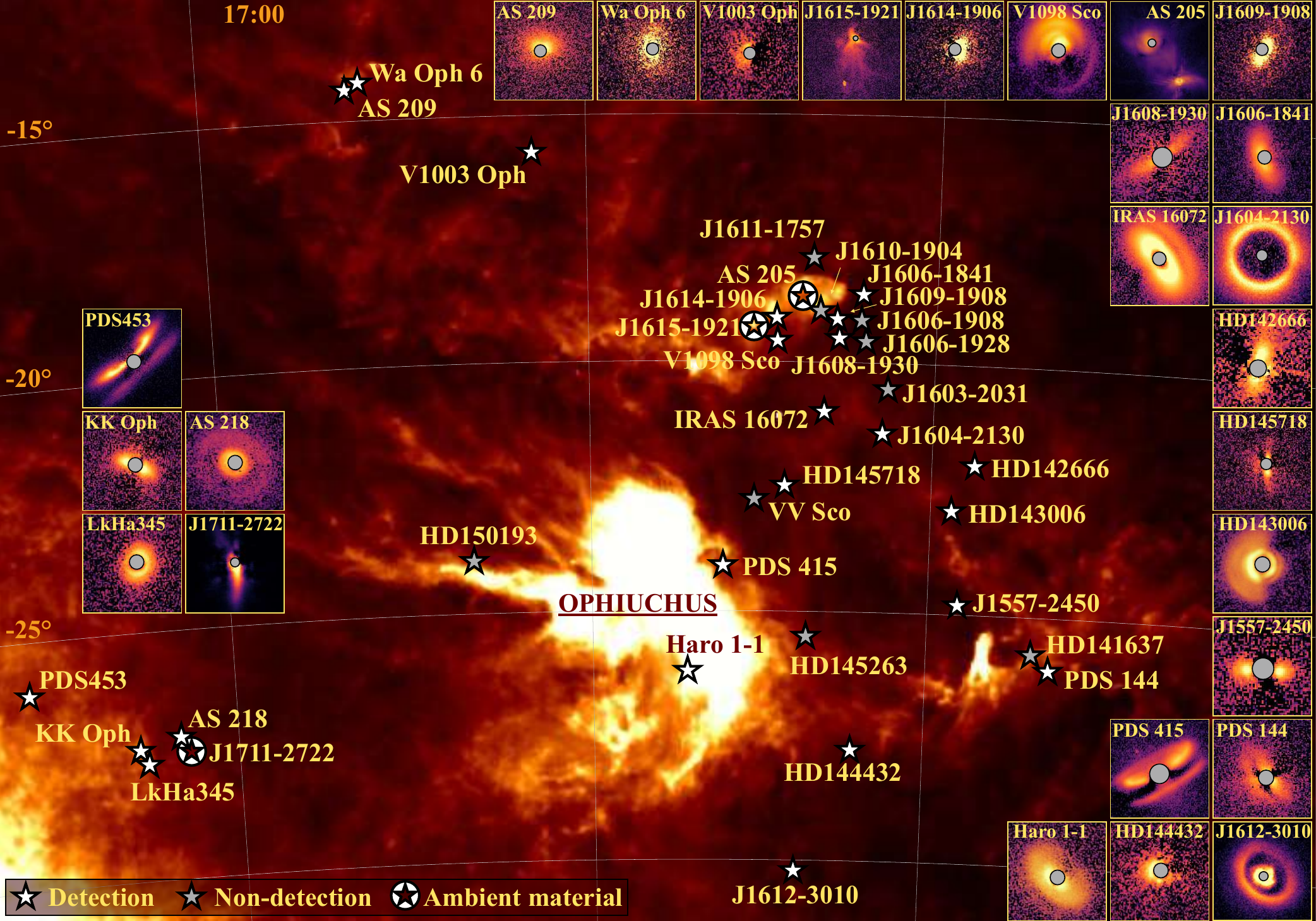}
    \caption{Spatial distribution of sample in {the Sco-Cen association. Sources in Upper and Lower Centaurus (as well as the outlying HD169142 and HD152404) are in the top map while sources in Upper Scorpius are in the bottom map. No source in Ophiuchus or Lupus is shown here (see Fig.\,\ref{fig:map_2}). All sources from Centaurus as well as peripheral sources from Upper Scorpius host bright disks}. The background image is from the Improved Reprocessing of the IRAS Survey (IRIS) at 100 $\mu$m (Band 4). The NIR images shown in the inset panels have spatial and flux scales optimized for better representation. Only detections are shown.}
    \label{fig:map_1}
\end{figure*}

The stellar variability turns out to be the quantity changing most dramatically between sources with ambient signal (median 0.9 mag) and without it (0.4 mag). To put it in another way, as much as 40\% of highly variable stars (more than 1.0 mag) show ambient signal, while this fraction is only 6\% for lowly variables (less than 0.3 mag). This trend is visible from the diagram of Fig.\,\ref{fig:ambient}, where it is clear that the detectable ambient material is also related to the NIR excess, and that overall NIR excess and stellar variability mildly correlate. A similar trend is present for the mass accretion rate despite the lower number of sources with this measurement available. In fact, the median accretion normalized by the stellar mass (Sect.\,\ref{sec:sample_variability}) of sources with and without ambient signal is $2\cdot10^{-8}$ and $5\cdot10^{-9}$ $\rm M_\odot yr^{-1}$, respectively. All this highlights a direct connection between the ambient material, its possible accretion onto the disk, and the subsequent accretion onto the star (see Sect.\,\ref{sec:discussion_infall}).

The diagram of Fig.\,\ref{fig:ambient} also shows a good partitioning of disks with rings and spirals (see also Sect.\,\ref{sec:analysis_morphology_substructures}). Sources with disk spirals have a large NIR excess and stellar variability, and are typically associated with ambient signal. In fact, as much as 52\% of the detected disks in sources with ambient signal show spirals. This fraction is very similar for shadows (48\%) while it drops to 0\% for rings. Our census reveals indeed no disk with rings in sources exhibiting ambient signal. This sharp difference clearly demonstrates that the morphology of disks is impacted by their interaction with the environment, and it is further discussed in Sect.\,\ref{sec:discussion_infall}.

\begin{figure*}
        \centering
    \includegraphics[width=13cm]{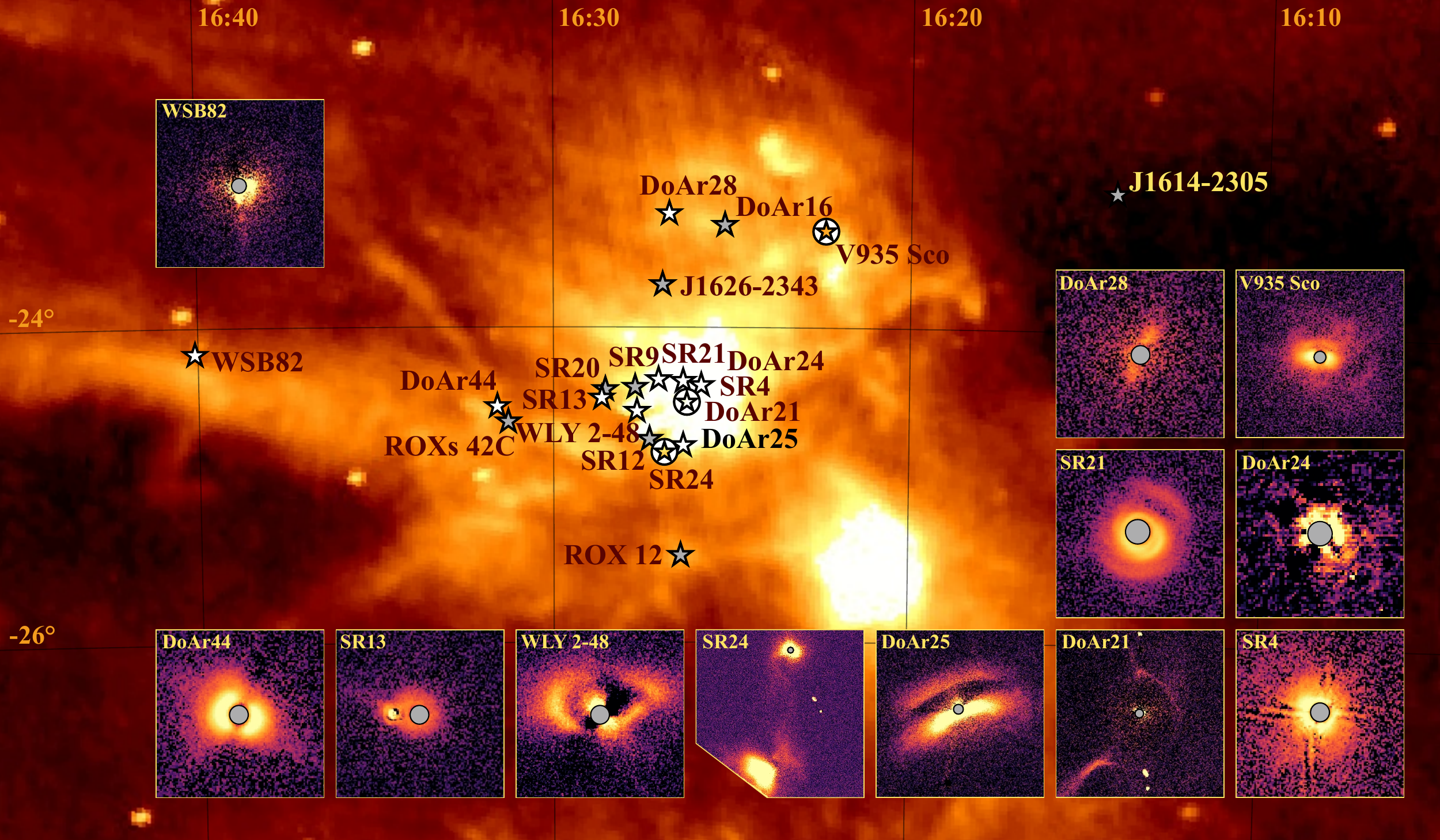}
    \includegraphics[width=13cm]{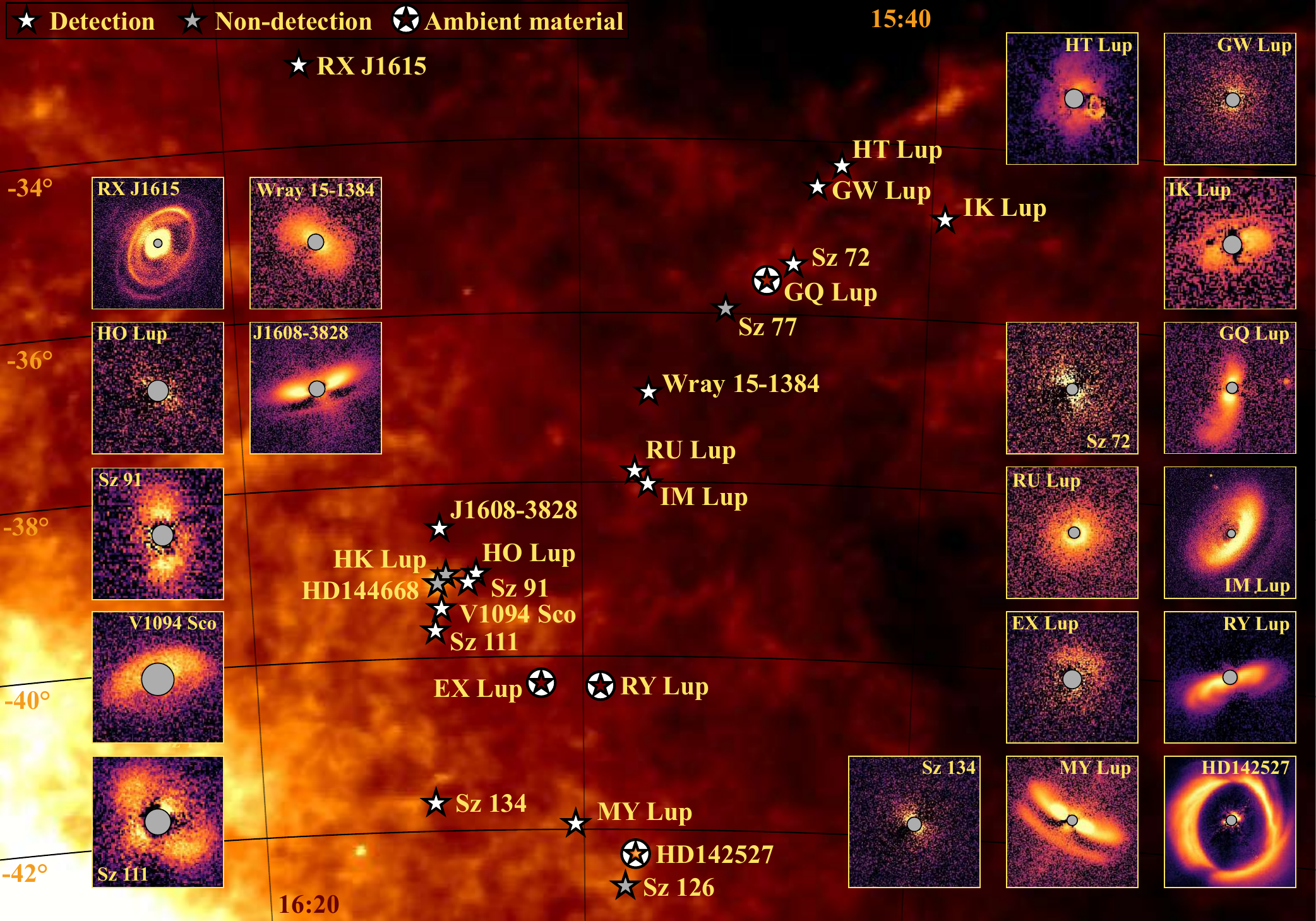}
    \caption{Same as Fig.\,\ref{fig:map_1} for Ophiuchus (top) and Lupus (bottom).}
    \label{fig:map_2}
\end{figure*}

\begin{figure*}
        \centering
    \includegraphics[width=13cm]{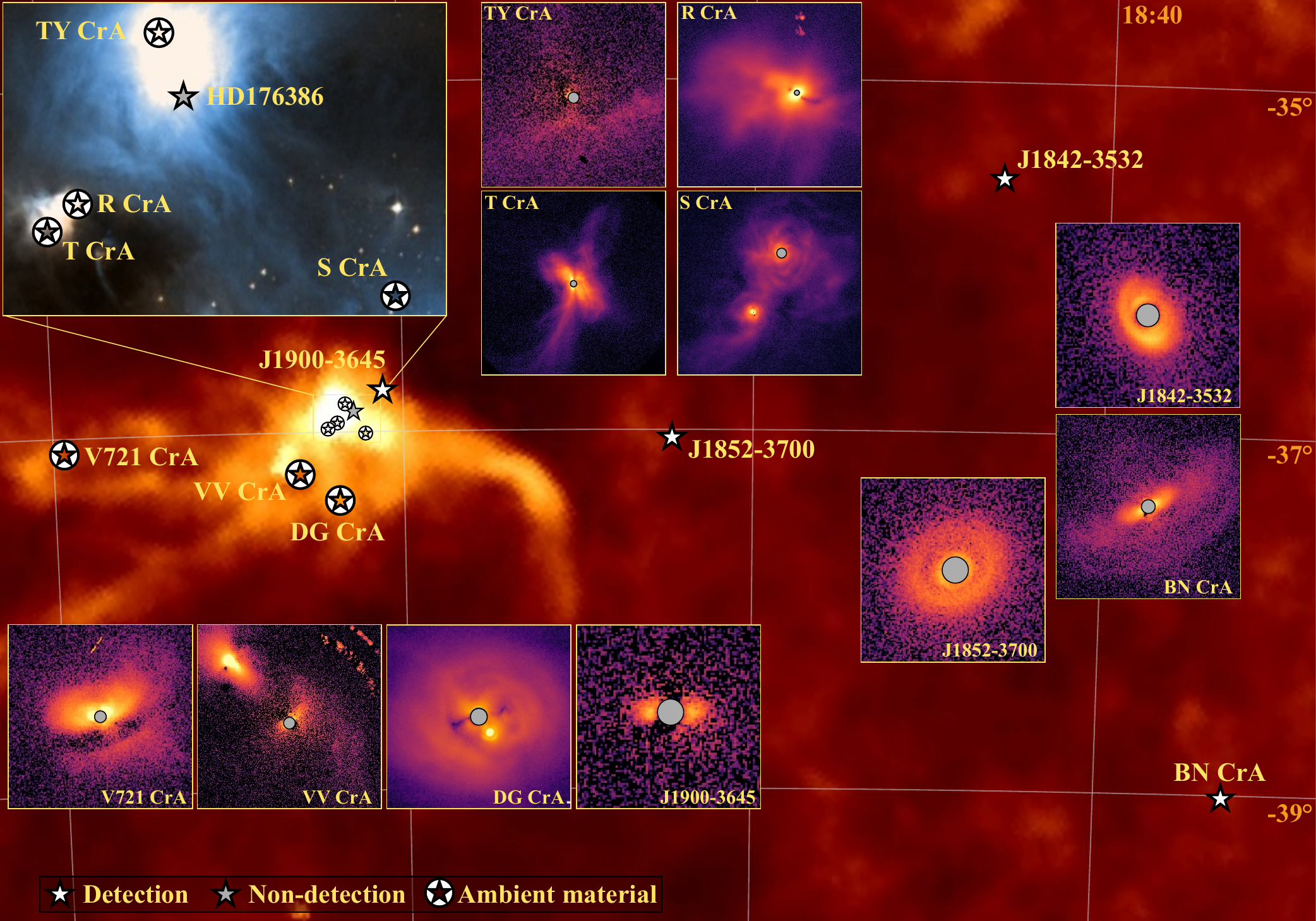}
    \caption{Same as Fig.\,\ref{fig:map_1} for Corona Australis. The inset image of the Coronet Cluster is from the Digitized Sky Survey (DSS) from the Palomar and UK Schmidt telescopes in the Blue band (0.47 $\mu$m).}
    \label{fig:map_3}
\end{figure*}

\subsection{Regional morphology} \label{sec:analysis_maps}

A first attempt to relate local disk properties from high-contrast imaging with the source location within its natal cloud was presented by \citet{Garufi2024}. They reported brighter disks in the peripheral regions of Taurus, along with a higher incidence of detectable ambient material. In contrast, \citet{Ginski2024} found that in Chamaeleon any such correlations were less evident, and that the disk appearance in the region is primarily influenced by the presence of stellar companions, which account for a large fraction (55\%) of the available sample. In this section, we scrutinize the spatial distribution of local properties in the other nearby star-forming regions.

The map of the large Sco-Cen association is shown in Fig.\,\ref{fig:map_1}. All disks in Upper Cen and Lower Cen are detected and often exhibit peculiar substructures such as a cavity with known protoplanets (PDS 70), spirals (HD135344B, HD100453), and shadows (HD139614, Wray 15-788). No strong spatial segregation in disk appearance is evident in Upper Sco. However, the non-detection rate (which is 28\% across the entire region) rises to 45\% in the densely populated areas around $\nu$ Sco and $\beta$ Sco (right above 20\degree\ declination). In contrast, all disks around more peripheral sources are detected, including Wa Oph 6, AS 209, and V1003 Oph (off-cloud to the NE), HD144432 and J1612-3010 (south of $\pi$ Sco), five stars near the Pipe Nebula (SE), as well as HD152404 and HD169142 (further SE, close to the galactic plane). This pattern resembles that observed in Taurus, where primarily bright disks lie toward the outskirts of the region. Unlike Taurus, however, the fraction of sources showing ambient emission remains low (8\%) and does not increase outward.

A closer look to the more compact Ophiuchus and Lupus regions is shown in Fig.\,\ref{fig:map_2}. Eleven of our 20 Ophiuchus targets are projected against the L1688 dark cloud (the bright central core at 100 $\mu$m in the figure). Eight of these sources are detected, and two show ambient emission. The fraction of non-detections (45\%) is comparable with the aforementioned dense clustering of Upper Sco. No significant spatial segregation of disk morphology is visible. The fraction of sources with ambient emission (15\%) is consistent with the  median of the whole sample, despite the strong presence of diffuse dust in the cloud, as also reflected in the large median $A_{\rm V}$ of our sample (2.9 mag). As in Ophiuchus, the spatial distribution of sources in Lupus shows no clear link with disk brightness or presence of ambient material. Exceptionally bright disks (e.g., IM Lup, J1608-3828, and HD 142527), non-detections, and sources with ambient emission are all found across the Lupus I–II (NW), Lupus III (East), and Lupus IV (South) clouds. As commented in Sect.\,\ref{sec:analysis_brightness}, the overall disk brightness is higher than in the rest of the sample (0.6\% vs 0.2\%), while the fraction of sources with ambient emission remains comparable (15\%). 

Finally, the map of Corona Australis is shown in Fig.\,\ref{fig:map_3}. Most available sources in this region lie within the dense Coronet Cluster surrounding R CrA. All cluster members (except the undetected HD176386) show clear evidence of ambient material, highlighting the strong interaction between the stars and their local environment. In contrast, the three sources located outside the cluster all host bright and prominent disks.

\section{Discussion} \label{sec:discussion}

The analysis of Sect.\ref{sec:analysis} led to several results regarding ($i$) the local evolution of planet-forming disks, ($ii$) the demographics of different disk populations, and ($iii$) the interaction of disks with the environment. These three topics are discussed in this section. It should be kept in mind that the our considerations mainly hold for stars with mass between 0.7 M$_\odot$ and 3 M$_\odot$ as lower-mass stars (unless very young) are too faint to be observed with the current high-contrast imaging instruments, while higher-mass stars are too far to be meaningfully resolved by 8-m telescopes.  

\subsection{Emergence of sub-structures and longevity of disks} \label{sec:discussion_longevity}
Our understanding of the planet-forming disk evolution is dependent on the determination of the individual stellar age, that is generally considered a challenging task for stars younger than \mbox{20 Myr} \citep[see review by][]{Soderblom2014}. The advent of \textit{Gaia} and of the last-generation high-contrast imaging helped reduce the observational uncertainty. The latter is particularly important to determine the presence of previously unresolved stellar companions at sub-arcsec scales \citep[e.g.,][]{Biller2012, Mesa2019} and of significantly inclined circumstellar disks ($\gtrsim60\degree$) that can lead to an underestimation of the stellar brightness and thus to an overestimation of the stellar age \citep[and this work]{Garufi2018}. With the notable exception of a few intermediate-mass stars (UX Tau, MWC480, and HD142527), all unflagged sources from this work (Sects.\,\ref{sec:sample_properties} and \ref{sec:sample_stars}) are in line with the age of the respective region. This emphasizes the importance to minimize the observational uncertainties while employing isochrones to constrain the age.

The compilation of this sample provides a significant number of long-living disks that are of particular importance to constrain the disk evolution. As many as 24 unflagged sources older than 8 Myr are in the sample. This work showed that practically all these disks are bright in scattered light (when closer than \mbox{200 pc}), indicating a disk inner region that is significantly depleted in dust. In turn, this supports the widely accepted idea that a strong pressure bump capable of retaining disk material at large separations is needed for a disk to live significantly longer than the typical lifetime of 2--3 Myr \citep[e.g.,][]{Mamajek2009}. The high mass of these old disks (Sect.\,\ref{sec:sample_disks}) can be explained by a survivor bias, as the dissipation of lighter disks unable to open cavities results in an increase of median mass \citep[see][for the photoevaporation case]{Malanga2025}. The low incidence of external companions in long-living disks (Sect.\,\ref{sec:sample_multiplicity}) is consistent with the tidal truncation exerted by the external star on the circumprimary disk \citep[see review by][]{Cuello2025} that results in smaller disks \citep{Garufi2017, Akeson2019, Manara2019, Zurlo2020} and in a steeper density profile facilitating radial drift \citep{Rosotti2018, Cuello2019}. Our analysis, in particular, uncovers a stark effect only from companions closer than 300 au.

This work also proves that disk NIR cavities are very uncommon in single stars younger than 3 Myr, whereas several disk cavities around young multiple systems are found \citep[e.g.,][]{Kraus2020, Zurlo2021}. These stellar companions are likely carving the observed circum-multiple disk cavities \citep{Artymowicz1994, Ragusa2020} and altering the disk morphology with the formation of shadows and spirals \citep[e.g.][]{Keppler2020, Toci2024}. If unseen planets are instead responsible for cavities around single stars, their carving of small dust grains (and thus perhaps gas) may be a relatively slow process. This contrasts to the carving of cavities in larger grains that is suggested by the existence of ALMA disk cavities from disks younger than \mbox{3 Myr} \citep{Sheehan2017, Huang2020}, although even the fraction of ALMA cavities appears to increase after 2 Myr \citep{Vioque2025}. After the dust trap responsible for the observed NIR cavities is in place at $\sim$3 Myr, this would suffice to preserve the disk for up to 20 Myr (the age of the oldest disks from our sample) or even 30 Myr \citep{Silverberg2020, Long2025}.

Unlike cavities, disk rings and spirals are found at any disk age. Notwithstanding the limited statistics on such a small sub-sample, rings are typically faint in young disks. This is most likely related to the overall NIR faintness of these objects, and several examples of rings in very young disks are in fact revealed by ALMA \citep{Sheehan2018, SeguraCox2020}, although disk sub-structures in Class 0 and I objects appear infrequent \citep{Ohashi2023}. Contrary to rings, disk spirals in the NIR are generally bright at any disk time. This would intuitively advocate for some perturbations that affect the normal disk geometry and locally increases the disk vertical height even in faint disks. All in all, the sharpest difference in the source's properties between spirals and rings concerns stellar variability, presence of shadows, and ambient material, as is discussed in Sect.\,\ref{sec:discussion_infall}.

\subsection{Disk populations across regions and time}
\label{sec:discussion_regions}
The seven star-forming regions with an appreciable sub-sample of NIR images are rather diverse in size, age, and gaseous content. Taurus, Ophiuchus, Chamaeleon, and Lupus have a widely accepted age of 1--3 Myr (where the median age of our sub-samples is 1.3--2.1 Myr). The number of their members is coarsely comparable as it approximately spans from 500 (Taurus) to 150 (Lupus). All four regions host an approximate 50\% of Class II among their members (see Table \ref{table:region_fraction}). Nonetheless, the typical morphology of their disks is rather diverse. Lupus and Taurus disks are systematically detected (82\%--86\%) unlike Chamaeleon and Ophiuchus (55\%). Considering detections only, Lupus disks are more than five times brighter than Chamaeleon disks. All in all, the disk typical morphologies across regions is already captured by their SED (see NIR-FIR excess diagram in Fig.\,\ref{fig:unres_prop}), and thus the median appearance of the NIR images can somehow be predicted.

The key question is whether these differences are due to different evolutionary paths (possibly driven by the environment, see Sect.\,\ref{sec:discussion_infall}) or different evolutionary stages (despite the minor age difference). A possible explanation for the different typical nature of Taurus disks (full, faint) and Lupus disks (cleared, bright) is that the latter objects are slightly older, and the inside-out disk sculpting has significantly proceeded to form cavities that make the Lupus disks bright in the NIR. This in principle consistent with the abrupt increase in the median disk brightness between 1--2 Myr and 3--5 Myr (Fig.\,\ref{fig:contrast_trend}), which is an age interval roughly centered on the Lupus age ($\sim$3 Myr). Instead, the observed larger multiplicity of the Chamaeleon solar-like stars (Sect.\,\ref{sec:sample_multiplicity}) is likely responsible for their low detection rate \citep{Ginski2024}, though the origin of this larger multiplicity is currently unexplained. Finally, the elusive nature of the Ophiuchus disks can be explained by their low mass. In fact, the median Class II disk mass is lower than that of Taurus and Lupus \citep{Williams2019}, whereas the median mass of all objects \citep[including the abundant Class I and possibly extincted Class II,][]{Evans2009} is comparable \citep{Cieza2019}. Therefore, NIR imaging may not be probing a family of disks comparable to that of Taurus that, in Ophiuchus, are still embedded or heavily extincted. 

Corona Australis is slightly older than these four regions. Its age is actually debated since literature values span from less than 3 Myr \citep{Peterson2011} to 6--9 Myr \citep{Ratzenbock2023}. A recent work by \citet{Rigliaco2025} constrained an age of 5.2 Myr for CrA main \citep[in line with][]{Galli2020} although the authors caution that a lower median value is possible since several young sources are severely embedded. Our small sub-sample exhibit a median age of 3.8 Myr (3.1 Myr for sources in the probably younger Coronet cluster, see Sect.\,\ref{sec:analysis_maps}). The very high fraction of disk detections reported in this work (11 out of 12 targets) is clearly related to the abnormally high frequency of sources with ambient emission (more than twice than any other regions), and is further discussed in Sect.\,\ref{sec:discussion_infall}.

The large Upper Sco region ($\sim$2100 members) contains several clusters with varied ages spanning from 4 to 14 Myr \citep{Ratzenbock2023}. Therefore, the variety of ages shown by our sub-sample (from $\lesssim$1 to $\gtrsim$15 Myr -- with a median of 5.8 Myr) is likely real. In particular, Upper Sco is, together with Taurus \citep{Garufi2024}, the only region where some spatial segregation of disk morphology is visible from the map, being all 13 peripheral disks bright (Sect.\,\ref{sec:analysis_maps}). This is possibly explained by the complex star-formation history of the entire region \citep{MiretRoig2022}, where different ages and environmental conditions determine the current disk morphology in different sub-clusters. This would also explain why only the two large regions Upper Sco and Taurus show the aforementioned segregation, as opposite to more compact regions such as Ophiuchus with a more uniform history of star formation. 

Since Orion is the closest massive star-forming region and it harbors a few thousand members with very varied age \citep{Hillenbrand1997, DaRio2010}, similar spatial trends are likely also present therein. In addition, the impact of the UV field on the disk morphology is also important in this region \citep{Valegard2024}. However, in this work we showed the dramatic effect of the Orion distance (390 pc) on the appearance of NIR images (with up to 80\% of the flux lost when compared to nearby regions). In view of this, any direct comparison with the other regions is currently discouraged.

Finally, our sample contains few sources from other old star-forming regions such as the small (23 members), very close (100 pc) $\eta$ Cha \citep[7--9 Myr,][]{Roccatagliata2024}, and the oldest sub-region of the Sco-Cen association Lower and Upper Centaurus \citep[16 Myr,][]{Pecaut2016}. As is discussed in Sect.\,\ref{sec:discussion_longevity}, all our sources from such old regions are bright in scattered light, and represent the lone survivors of an old population of planet-forming disks. The most notable examples of this category are CQ Tau and MWC758 in the 14-Myr old HD35187-association near Taurus \citep{Luhman2023}, HD135344B, HD139614, and PDS 70 in the 16-Myr old Upper Cen \citep{Pecaut2016}, and V4046 Sgr in the 18-Myr old $\beta$ Pic moving group \citep{MiretRoig2020}.

\subsection{Impact of late infall on the disk structure} \label{sec:discussion_infall}
A most notable finding of this work is the large overall fraction of sources showing ambient signal (approximately 20\%). This result continues a trend that has been observed by many in both scattered light \citep[e.g.][]{Ginski2021, Huang2022, Garufi2024} and molecular line observations \citep[e.g.,][]{Akiyama2019, Yen2019, Yang2020, Speedie2024, Speedie2025}. This ambient signal likely probes infall of gas from the interstellar medium (ISM) on Class II sources, much later than the traditionally assumed funneling of material from
the collapsing envelope onto the protostar \citep{Terebey1984}. This has led to a revival of the debate about whether this late infall may have a substantial impact on the evolution of disk masses and angular momentum transport \citep[e.g.][]{Kuffmeier2023, Winter2024}. If infall proceeds at a sufficiently high rate, apart from adding mass \citep{Manara2018}, this could have numerous consequences for disk physics. For example, it may drive instabilities and dust structures \citep{Bae2015}, disk warping, misalignment and shadows \citep{Kuffmeier2021, Krieger2024}, or turbulence and outbursts \citep{Zhu2010}. This possibility is pertinent to the results of this work in three ways, that are the association of disk sub-structure with the detected ambient material, the different fraction of ambient material between regions, and the temporal evolution of the surviving disk population. 

The analysis of Sect.\,\ref{sec:analysis_ambient} provides the first statistical study of the impact of infall on the disk structure. From Fig.\,\ref{fig:ambient}, it is clear that the presence of ambient material is correlated with disk spirals, higher stellar variability and higher NIR excess. The most compelling evidence of the connection in question is that half of the disks encircled by ambient material shows spirals and shadows {in scattered light} while none of them show rings. {The NIR census presented here is, however, sensitive only to the structure of the disk upper layers as traced by scattered light, and ALMA surveys probing the disk midplane may therefore yield different results (see, e.g., HL Tau, where a ringed disk is enshrouded by ambient material, \citealt{Yen2019, Garufi2022a, Mullin2024}).} Material accreted from the ISM late in the disk lifetime has a random angular momentum vector \citep{Kuffmeier2024} and may therefore introduce both density waves and warping that induces spiral structures in scattered light \citep{Winter2025, Calcino2025}. The connection between ambient material and spiral structures is even clear in a few young sources (e.g., DR Tau and WW Cha). In some other cases, such as MWC 758 and CQ Tau, no ambient material is detected but the disk is suggested to be warped based on kinematic structure \citep{Winter2025} and it exhibits SO emission \citep{Zagaria2025}, which may originate at the shock front from the infalling material \citep{Garufi2022a, Speedie2025}.

\begin{table}[t]
\centering
\caption{Mean gas surface density, stellar variability, and incidence of ambient material.}
\begin{tabular}{lccc}
\hline
Region & $\langle \Sigma_{\rm gas} \rangle_{\rm YSO}$ & $\Delta{\rm V}$ & $f_{\rm ambient}$ \\
 & $( \rm M_\odot\ pc^{-2})$ & (mag) & \\
\hline
Taurus & 108.0 & 0.7 & 28\% \\
Ophiuchus & $105.4\pm41.0$ & 0.3 & 16\% \\
Corona Australis & $92.3\pm12.0$ & 1.0 & 58\% \\
Upper Scorpius & $85.2\pm22.0$ & 0.3 & 8\% \\
Chamaeleon & $84.1\pm10.6$ & 0.4 & 11\% \\
Lupus & $62.5\pm12.5$ & 0.5 & 13\% \\
\hline
\end{tabular}
\tablefoot{Columns are region, mean gas surface density $\Sigma_{\rm gas}$ per YSO, stellar variability, and fraction of sources with ambient emission. $\Sigma_{\rm gas}$ is based on the findings of \citet{Heiderman2010}, weighted by number of YSOs in each sub-region (a total of 148 in Tau, 290 in Oph, 41 in CrA, 117 in Cha, 10 in Upper Sco, 181 in Lup). Uncertainties are propagated from the reported $\Sigma_{\rm gas}$ errors where available (not reported for Taurus). Regions are sorted by $\Sigma_{\rm gas}$.}
\label{tab:yso_weighted_sigma}
\end{table}

Late infall may also drive some of the differences in morphology between regions discussed in Sect.\,\ref{sec:discussion_regions}. The rate at which material is captured onto the disk is sensitively dependent on both the local density and velocity structure of the ISM \citep{Bondi1944, Edgar2004}. Therefore, large-scale variations in the ISM properties between regions may induce variations of disk properties. For example, \citet{Gupta2023} showed that sources with large-scale gas structures are associated with nearby reflection nebulae. Also, \citet{Winter2024c} reported a correlation
between the stellar accretion rates and the local ISM density in Lupus. A correlation between ambient material and accretion rate is in fact also observed in our sample, though with large scatters (Sect.\,\ref{sec:analysis_ambient}). While a full study correlating disk properties across different regions with the local ISM is a substantial undertaking that warrants a devoted study, we can crudely compare the average surface density across regions with the fraction of ambient signal revealed in this work. To do this, we use the ISM surface density of local regions as reported by \citet{Heiderman2010}, aggregating the sub-regions into the larger regional definitions that we adopt in this work, weighted by the number of YSOs in each sub-region. The result of this calculation is reported in Table \ref{tab:yso_weighted_sigma}. While we lack statistical power in the sample size, we can see clearly that Taurus, Corona Australis and Ophiucus have the highest ISM surface density as well as the highest incidence of ambient material detected in scattered light. Corona Australis and Taurus also have the highest variability of the sample. This may be consistent with higher rates of infall across these regions, driving changes in the disk structure and perhaps transport of material inwards. Clearly this is suggestive rather than conclusive, but strongly motivates more focused studies establishing the per-star surface density versus disk property correlations.

The effect of infall in an evolutionary sense is more challenging to interpret quantitatively. It is clear that the incidence of ambient material decreases with time (Sect.\,\ref{sec:analysis_ambient}), as would be expected due to decorrelation ISM density and velocity structure with the star over time. However, as also expected, this is a stochastic trend as, for example, the Coronet cluster in the Corona Australis has the highest ambient material detection rate but the disks are relatively old ($3-5$ Myr, see Sect.\,\ref{sec:discussion_longevity}), although it retains a high ISM surface density (Table \ref{tab:yso_weighted_sigma}). This raises the possibility that the survival of older disks is partly determined by which systems underwent the latest substantial infall events. This would qualitatively fit with the higher prevalence of shadows among older disks (Sect.\,\ref{sec:analysis_morphology_shadows}). In the infall scenario, these disks would gain fresh material later in their life from material with angular momentum uncorrelated with the initial disk, producing a misaligned system. However, it is unclear if a primordial misalignment of the inner disk ($<$5 Myr) with a typical accretion rate could survive for the age of the older systems ($>$8 Myr) as, for example, a misaligned inner disk of mass $\rm 10^{-3}\,M_\odot$ and stellar accretion rate $\rm 10^{-9}\ M_\odot$~yr$^{-1}$ would disappear in barely 1 Myr. Therefore, some of the shadows and spirals observed in the oldest disks may not be due to late infall but rather to other environmental processes such as the interaction with a stellar companion like in the case of HD100453 \citep{Benisty2017, vanderPlas2019}.

\begin{figure*}
        \centering
    \includegraphics[width=17cm]{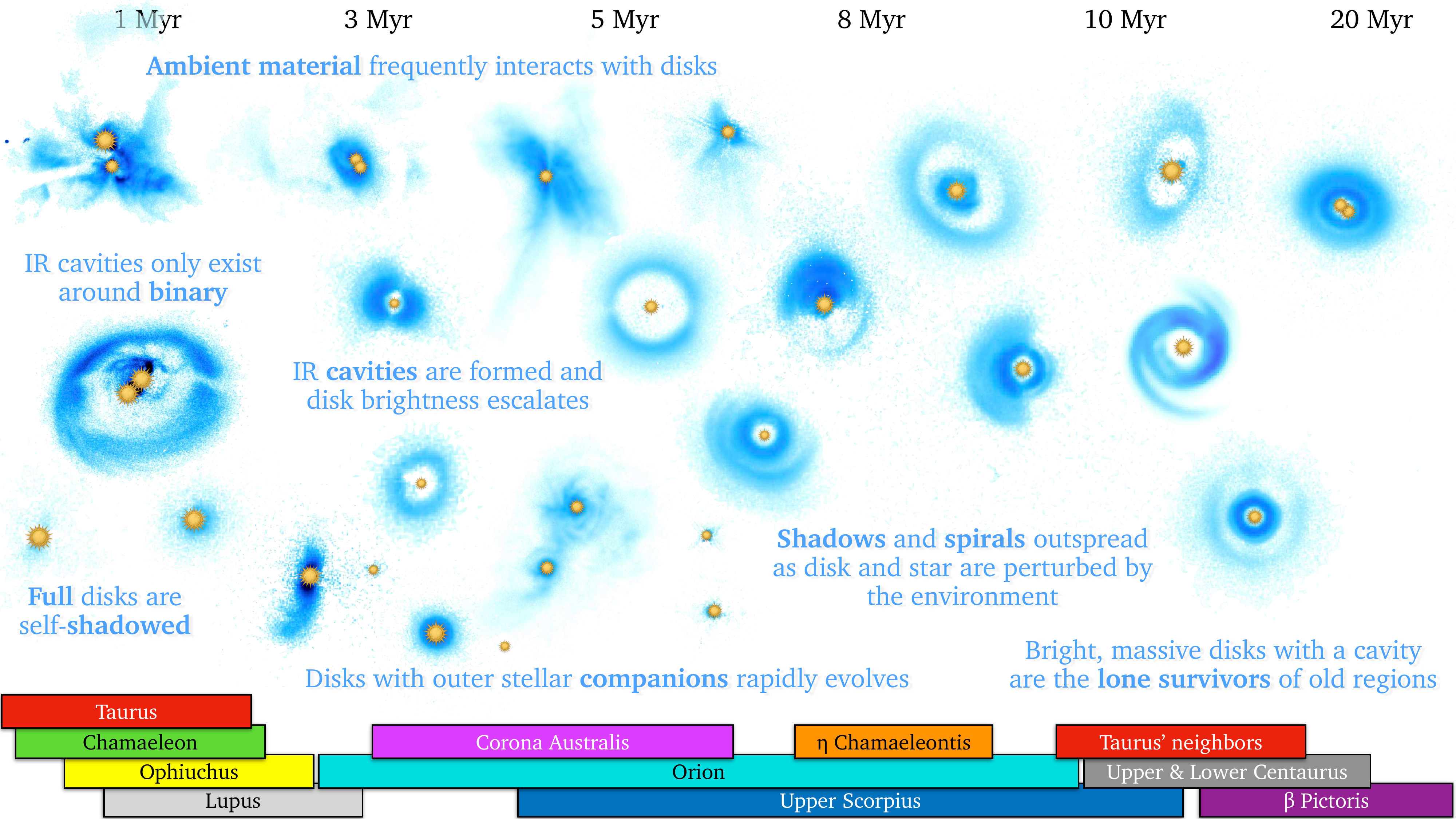}
    \caption{Visual synopsis of the census.}
    \label{fig:sinopsis}
\end{figure*}

\section{Summary and conclusions} \label{sec:conclusions}

Drawing on fifteen years of near-IR high-contrast imaging with advanced AO systems, we assembled a sample of 268 sources and carried out the largest infrared census of planet-forming disks to date. Several results concerning the evolution of disk and ambient material emerged from the census, as is summarized in the visual synopsis of Fig.\,\ref{fig:sinopsis} and below:

\begin{itemize}
    \item The available sample has a good level of completeness (50\%--90\%) for solar-mass stars and above ($M \rm>0.7\ M_\odot$) in nearby star-forming regions ($<200$ pc). Low-mass stars and further objects (including those in Orion) are much less characterized. 
    \item Individual stellar ages from isochrones are in line with the age of the hosting region when critical cases are isolated. These include stars with a very inclined disk that partly occult the stellar flux.
    \item Disks from different -- even coeval -- regions appear generally different. These differences are also visible from their SED. Lupus disks are five times brighter than Chamaeleon disks. The disk detection rate in Taurus is similar to Lupus but disks therein are three times fainter.
    \item The average disk brightness increases with time with an abrupt rise from 1--2 Myr to 3--5 Myr. Disks older than 8 Myr are systematically bright. 
    \item No NIR disk cavity is detected around single stars younger than 2.5--3 Myr. This explains the faint nature of young disks. 
    \item A handful of compact stellar systems host giant disks with a large cavity and multiple spirals. Instead, disks are rarely detected when an outer companion within 300 au is present. Old disks in this configuration are very rare suggesting that their evolution is faster.  
    \item Sub-structures (cavity, spirals, rings) are detected in practically all bright, non-inclined disks.
    \item Local shadows are seen in 15\%--20\% of disks. This fraction increases with time peaking to 35\% in the \mbox{4--10 Myr-old} \mbox{Upper Sco}. Their presence is related to a higher stellar variability and NIR excess.
    \item Disk spirals and rings naturally separate stars with high and low NIR excess (25\% vs 10\%) and variability (0.6 vs 0.3 mag).
    \item Ambient material is detected in 15\%--20\% of all sources. This fraction decreases with time but is maximized (60\%) in the 3--6 Myr-old Corona Australis.
    \item Like spirals and rings, sources with and without ambient material naturally separate high and low NIR excess (25\% vs 10\%) and stellar variability (0.9 vs 0.3 mag), as well as mass accretion rate (by a factor three). Remarkably, half of the disks with ambient material show spirals while none of them show rings.
\end{itemize}

The final picture that emerges is that the evolution of planet-forming disks is driven by both inside and outside effects. From inside, the formation of a cavity for small dust grains (and possibly gas) which is occurring not earlier than 2--3 Myr represents a fork for the future disk evolution. The fraction of planet-forming disks in clusters decreases from more than 50\% to approximately 20\% in a time range of a few Myr around an age of 3 Myr. The disks surviving this selection would be those capable of opening a cavity. The high longevity of several disks from our census (more than 20 older than 8 Myr) shows that this geometry could then sustain the disk existence for another 10–15 Myr. While this work does not aspire to explain the origin of disk cavities, in the current framework this would be the formation of giant planets in the inner 10–20 au. As of today however, only PDS 70 among the high-longevity disks is known to host planets. 

From outside, the disk morphology is influenced by both external stellar companions and ambient material. The more than 20\% of sources younger than 3 Myr that show ambient signal in the current snapshot of time clearly indicates that this is an essential aspect of the disk evolution. Although the overall fraction of these sources decreases with time, disk-ambient interaction could also be revived or sustained for {a} long time in disks embedded in high gas surface-density environment, as for example in Corona Australis ($\gtrsim$3 Myr). This work provides further evidence to the substantial accretion of ambient material on the disks. In particular, we propose that the disk NIR spirals would originate from the perturbation induced by external accretion (or interaction with other stars in a few cases). This also leaves an imprint on the disk by inducing warps (traced by shadows and increased near-IR excess) and on the star by augmenting stellar variability and mass accretion rate. This scenario would also explain the long-known analogy between spirals, shadows, and near-IR excess. 

Therefore, the varied disk morphology observed in different star-forming regions can also be explained by both inside and outside mechanisms. The striking diversity of the quasi-coeval Ophiuchus, Chamaeleon, Taurus, and Lupus is partly explained by their environment. Ophiuchus disks are more embedded and the population of bare Class II disks that is targeted in the NIR is less massive and thus more elusive. The Chamaeleon disks are impacted by the high stellar multiplicity. Instead, the difference between Taurus (faint disks) and Lupus (bright disks) is more intriguing as the two regions are similar in morphology, population, extinction, dimension, and age. Since the Lupus age is very close in age to the aforementioned fork of disk evolution at \mbox{3 Myr}, one suggestive explanation is that we probe the rapid morphological changes driven by the disk cavity opening occurring in Lupus. Finally, increasingly old regions probe increasingly biased disk samples since the survivor population is dominated by exceptionally bright and extended disks, to the point that only objects like MWC758, HD135344B, and V4046 Sgr are found in 10--20 Myr-old regions.  

Taken together, the emerging picture may be one of disk diversity and evolution driven by the connection between the disk and a stochastically evolving local ISM via late infall and interaction with other stars, as well as by the interaction with giant planets capable of hampering maturation and producing a population of high-longevity disks. The future development of the current NIR census is entrusted to the Extremely Large \mbox{Telescope} since its observations will enable the sample expansion toward lower-mass stars, smaller planet-forming disks, and further star-forming regions.

\begin{acknowledgements}
      The research activities described in this paper were carried out with contribution of the Next Generation EU funds within the National Recovery and Resilience Plan (PNRR), Mission 4 - Education and Research, Component 2 - From Research to Business (M4C2), Investment Line 3.1 - Strengthening and creation of Research Infrastructures, Project IR0000034 – “STILES - Strengthening the Italian Leadership in ELT and SKA”. CFM is funded by the European Union (ERC, WANDA, 101039452). Views and opinions expressed are however those of the author(s) only and do not necessarily reflect those of the European Union or the European Research Council Executive Agency. Neither the European Union nor the granting authority can be held responsible for them. This research has made use of the VizieR catalogue access tool, CDS, Strasbourg, France (DOI: 10.26093/cds/vizier). The original description of the VizieR service was published in \citet{Ochsenbein2000}. This work presents results from the European Space Agency (ESA) space mission \textit{Gaia}.
\end{acknowledgements}


\bibliographystyle{aa} 
\bibliography{Reference} 

\appendix

\section{Sample properties and new images} \label{appendix:sample}

In Tables \ref{tab:sample_r1_g3}, \ref{tab:sample_r2_g3}, \ref{tab:sample_r3_g3}, \ref{tab:sample_r4_g3}, and \ref{tab:sample_r5_g3}, we list the NIR-image properties of the sample explored in this work. As described in Sects.\,\ref{sec:analysis_scrutiny} and \ref{sec:analysis_morphology}, we {label} any disk feature visible from the high-contrast image as cavity, spiral(s), ring(s), back side, broad shadow(s), for narrow shadow(s). Similarly, as described in \ref{sec:analysis_ambient}, we {label} any ambient feature as envelope, binary interaction, and ambient interaction with the environment. The complete table with all stellar and disk properties explored in this work is available online.

In the following, we present the original SPHERE images. {All detections are shown in Fig.\,\ref{fig:new_sources_gallery}.} As mentioned in Sect.\,\ref{sec:analysis_new}, 37 are from DESTINYS (none of which with prior publications), 9 from the young intermediate-mass star sample targeted by program 0111.C-0369 (of which 7 with no prior publications), and 7 from other programs (of which 4 with no prior publications).

\subsection{DESTINYS}

A large fraction of the 37 DESTINYS sources with no prior publications (11) is located in Upper Sco. The other well represented regions are Chamaeleon (7 -- all in Cha II since \citealt{Ginski2024} reported all Cha I sources) and Lupus (6). Twelve images show no detection of polarized light. These sources are \mbox{IRAS 12535-7623}, 2MASS J13005622-7654021, 2MASS J16265280-2343127, Hn 24, Hn 23, EP Cha, \mbox{Hen 3-854}, HD141637, Sz 77, Sz 126, EM\mbox{*} SR23, and 2MASS J17285631-2710031. {Yet, the first four sources of the list show a stellar companion that is shown in Fig.\,\ref{fig:new_sources_companions}.}

\begin{figure*}
    \centering
    \includegraphics[width=0.95\linewidth]{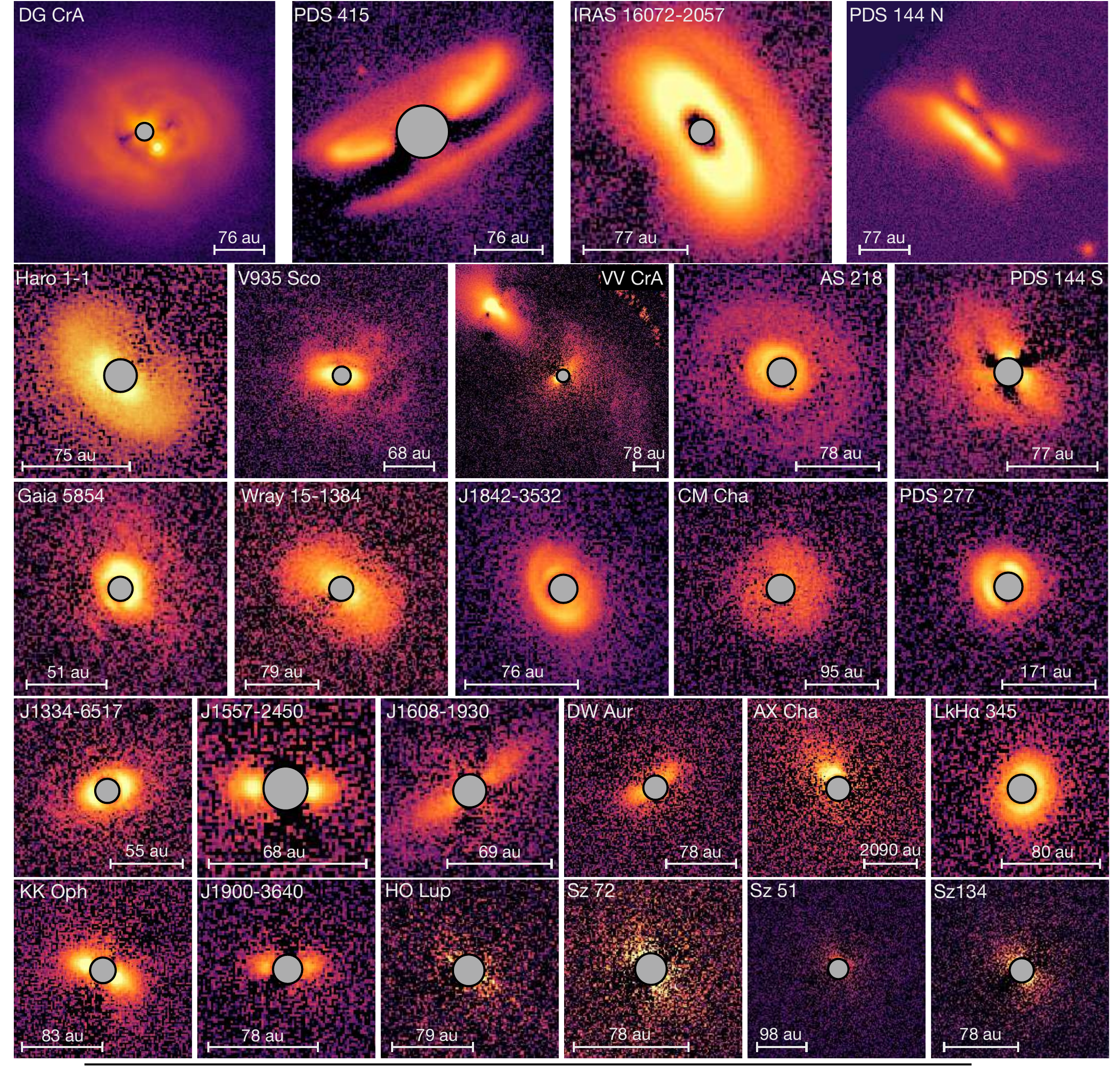}
    \includegraphics[width=0.95\linewidth]{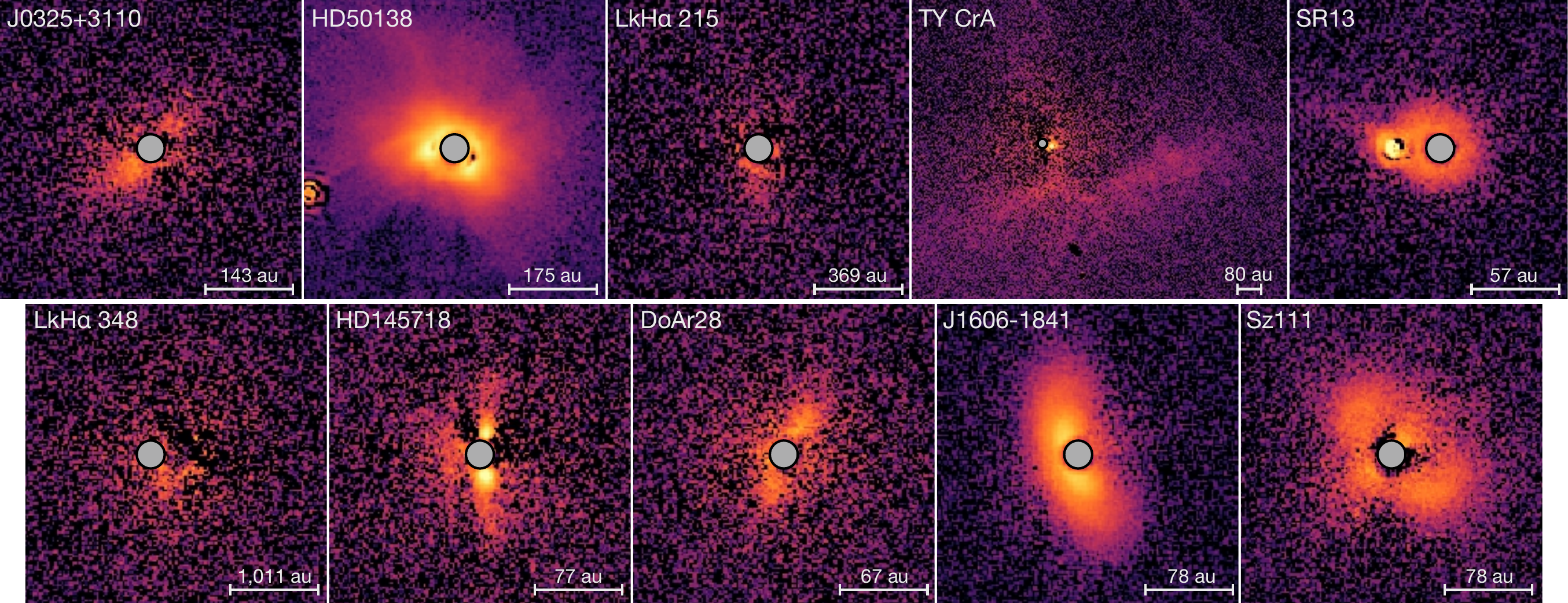}
    \caption{Gallery of the original images. The DESTINYS targets are gathered in the first five rows above the horizontal line. For each object, the $Q_\phi$ image is shown. The flux scale is arbitrary and different from source to source. The ruler indicates an angular size of 0.5\arcsec\ corresponding to the spatial scale indicated in each panel. }
    \label{fig:new_sources_gallery}
\end{figure*}

All 25 {polarimetric} detections are shown in Fig.\,\ref{fig:new_sources_gallery}. The appearance of the $Q_\phi$ image varies from bare detections to a few prominent detections that are among the most spectacular ever obtained from near-IR high-contrast imaging. {Below, we discuss each object individually, following the ordering in Fig.\,\ref{fig:new_sources_gallery}, which broadly reflects decreasing disk brightness}. \\

\textit{DG CrA}. This image shows one of the most prominent detection of the DESTINYS sample. A stellar companion is detected from the intensity image 0.22\arcsec\ south of the primary. A highly sub-structured disk is detected out to 1\arcsec\ from the star. Beyond this, the polarimetric image reveals several spiral arms with complex and varied morphologies extending out to 2.5\arcsec. One arm in particular, located to NE, stands out. It appears to connect the primary star, or possibly the secondary, to the surrounding environment in a relatively straight path, contrasting with the more tightly wound structure of all other arms. \\
\textit{PDS 415}. This is a showcase of SPHERE detections. The image shows a very inclined disk. Here, the disk upper rim seems to almost occult the star (since the disk is almost seen edge-on), and the disk back side is almost as bright as the front side. Interestingly, the northern part of the image reveals an asymmetric arm that is intuitively interpreted as a vertically high spiral arm spanning significantly more than 90\degree\ around the star. Thus, this is probably the side view of several spiral arms detected in more face-on disks \citep[e.g.,][]{Garufi2013, Benisty2015}. \\ 
\textit{IRAS 16072-2057}. An extremely clean detection of a moderately inclined disk is obtained from the image. An inner cavity of 25 au and the disk back side are also clearly resolved. The disk is so bright (the fifth brightest of the entire sample) that it is visible directly from the intensity image on top of the stellar PSF. \\
\textit{PDS 144 S}. The bright, extended disk (at least 90 au) is promptly detected from the image. Clearly negative signal is detected in the perpendicular direction. This is the result of the divergence for the Q and U images from the typical butterfly pattern generated by perfectly centro-symmetric scattering. A very inclined disk (more than 70\degree) is a possible geometry for such a divergence \citep[see e.g.,][]{Canovas2015}. The SPHERE image also reveals the prominent detection of the disk of PDS 144 N (shown in Fig.\,\ref{fig:new_sources_gallery} but not included in the sample in view of the considerations of Sect.\,\ref{sec:sample_publ}). This star lies 5.3\arcsec\ NE of the southern component, and hosts a perfectly edge-on disk known from Keck observations \citep{Perrin2006}. \\ 
\textit{Haro 1-1}. A moderately inclined disk is detected out to 70 au. A rather abrupt decline of disk flux occurs outward of 25 au. \\
\textit{V935 Sco}. This image shows both a disk and ambient contribution. The disk is clearly detected and it appears inclined, with a P.A.\ of 80\degree, and an apparent extent of 50 au. The disk back side is also visible. On top of this, some very asymmetric and diffuse material is detected in scattered light out to 150 au. \\
\textit{VV CrA}. Very faint polarimetric signal is detected along SE-NW direction pointing to an inclined disk with an approximate P.A.\ of 130\degree. A stellar companion is detected 2.1\arcsec\ NE of the primary star {(see Fig.\,\ref{fig:new_sources_companions})}. \\
\textit{EM\mbox{*} AS218}. The disk is clearly detected out to 80 au. A large gap separates the inner disk close to the coronagraph and an outer component that appears moderately inclined. \\
\textit{Gaia DR2 5854897321965963264}. Some signal is clearly visible around the coronagraph out to 0.2\arcsec. However, some faint signal is also detected out to 0.6\arcsec, and in particular to West. This outer signal suggests a relatively inclined with a P.A.\ of 10\degree--20\degree.  \\
\textit{Wray 15-1384}. A clear, moderately inclined disk is detected out to 80 au. A hint of disk backside is visible to NW. \\
\textit{RX J1842.9-3532}. A relatively small disk (0.35\arcsec) is clearly detected. It does however show several spiral arms with the most prominent one lying to NE. \\
\textit{CM Cha}. A faint but relatively extended disk (out to 0.4\arcsec) is visible from the image. The disk appears close to face-on {and shows evidence of shadowing on the eastern side}. \\
\textit{PDS 277}. A compact, rather bright disk is detected. It shows some asymmetric substructures that can be interpreted as spirals arms, although a broken ring is still a viable interpretation. \\
\textit{2MASS J13342461-6517473}. A disk with a P.A.\ of 100\degree\ is clearly detected around the coronagraph. A close inspection reveals the presence of a tenuous ring at 0.4\arcsec\ that suggests an inclined geometry for the disk. \\
\textit{2MASS J15575444-2450424}. Bright, compact signal is detected along the East-West direction. This suggests the existence of a very inclined disk seen horizontally.  \\
\textit{2MASS J16082324-1930009}. A very inclined, extended (80 au) disk is clearly detected. Despite the high disk inclination, the images show little divergence from centro-symmetric scattering. \\
\textit{DW Aur}. A clearly inclined disk with an approximate P.A.\ of 120\degree\ is visible. A tentative detection of the disk back side is visible to the SW. \\
\textit{AX Cha}. Polarized signal is detected in narrow wedge toward North only. Based on the extremely large distance to the source ($\sim$4200 pc), this signal would come from separations as large as 2,500 au from the central star. \\
\textit{EM\mbox{*} LkH$\alpha$ 345}. A faint polarimetric signal is detected only out to 0.2\arcsec. The emission is slightly more extended to SW than to NE suggesting a disk P.A.\ of approximately 160\degree\ with the near side lying to NE. \\
\textit{V* KK Oph}. A very inclined disk with a P.A.\ of 70\degree\ is visible from the image. A bright companion is detected from the intensity image at 1.6\arcsec\ SW of the star {(outside of the field of view of Fig.\,\ref{fig:new_sources_gallery} but visible from Fig.\,\ref{fig:new_sources_companions})}.  \\
\textit{2MASS J19005804-3645048}. A tiny disk (0.2\arcsec) is detected with a P.A.\ of 90\degree. The signal is only detected along the E-W axis indicating that the disk is very inclined. \\
\textit{V* HO Lup A}. This image only shows a marginal polarimetric detection at P.A.\ of 45\degree. A bright companion is visible at 1.54\arcsec\ {(outside of the field of view of Fig.\,\ref{fig:new_sources_gallery} but visible from Fig.\,\ref{fig:new_sources_companions})}. \\
\textit{Sz 72}. Positive polarized signal is detected to North and South of the star. {Similarly to the case of PDS 144 S}, some negative signal is also detected because of the very inclined nature of the disk. \\
\textit{Sz 51}. Some tenuous signal is detected around the coronagraph and seems to extend both to North and South. {However, the signal is too tenuous to enable any characterization of the disk geometry.} \\
\textit{Sz 134}. The disk is only marginally detected around the coronagraph. {This case is very similar to that of Sz 72 where a deviation from centro-symmetric scattering induced by a very inclined disk is possible.} \\

\subsection{Young intermediate-mass stars}
\begin{figure*}
    \centering
    \includegraphics[width=0.95\linewidth]{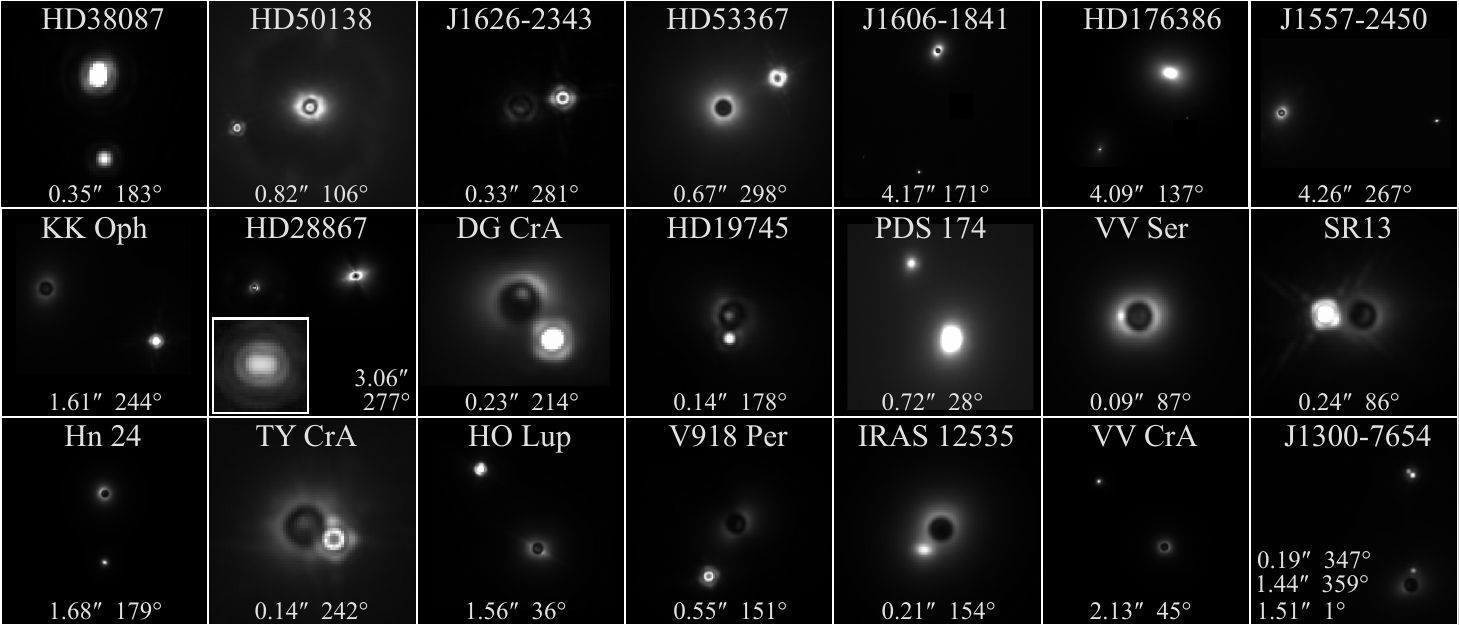}
    \caption{{Gallery of the original images with clear evidence of stellar companions. For each object, the $I$ image is shown. The flux scale is arbitrary and different from source to source. Separation and position angle (East of North) of all companions are given to the bottom of each image.}}
    \label{fig:new_sources_companions}
\end{figure*}

Nine stars were observed in the program 0111.C-0369 (\mbox{PI: Vioque}). Two of them (HD158643 -- or 51 Oph -- and HD50138) were observed with GPI by \citet{Rich2022}. No NIR image of the other seven sources from Vioque’s program has been reported. Here we describe the appearance of all SPHERE images from the program. The polarimetric detections are show in Fig.\,\ref{fig:new_sources_gallery}. \\
\textit{2MASS J03254982+3110237}. This polarimetric image shows a clear detection. A faint disk is visible, and it appears rather inclined (with an approximate P.A. of 110\degree\ East of North). \\
\textit{HD50138}. This star hosts a companion {(see Fig.\,\ref{fig:new_sources_companions})} at 0.82\arcsec\ (288 au) that is clearly detected from the intensity image. The $Q_\phi$ image shows several bright, extended features across multiple directions (see Fig.\,\ref{fig:new_sources_gallery}). To a first glance, the extended emission resembles the hourglass shape of outflow cavities. To larger separations however, all features appear highly intricate and asymmetric, thus suggesting some complex interaction with the environment rather than the classical dissipating envelope. \\
\textit{HD19745}. The intensity image shows a stellar companion at 0.14\arcsec\ corresponding to 56 au {(see Fig.\,\ref{fig:new_sources_companions})}. From the available non-coronagraphic images, we constrain the flux ratio to be around 4\% in the H band. The polarimetric images do not show any signal. Since the IR excess from the SED starts at 10 $\mu$m, this source is at the boundary of our sample with Class III (see Sect.\,\ref{sec:sample_publ}). \\
\textit{V918 Per}. This source hosts a stellar companion at 0.55\arcsec\ (168 au) that is detected from the intensity image, and is approximately one third fainter than the primary star in the H band. No polarized signal is detected. \\
\textit{HD28867}. This source is a bright X-ray source known to have a companion at 3\arcsec\ of a similar spectral type. \citet{Walter2003} concluded that the E component emits X-ray, has NIR excess, and hosts a companion within 14 au (unresolved in their image). From our image, we see an elongated inner PSF {(see inset image in Fig.\,\ref{fig:new_sources_companions})} that would point to a binary with separation of approximately 0.06\arcsec\ (10 au). No polarimetric signal is detected. \\
\textit{VV Ser}. The intensity image shows a companion at the edge of the coronagraph (at 0.09\arcsec\ or 37 au, {see Fig.\,\ref{fig:new_sources_companions}}). This is also visible from auxiliary non-coronagraphic images where it can be constrained that the flux ratio in the H band is smaller than 10\%. The polarimetric images are affected by the presence of the companion. Some signal may be visible from the Q and U images to the South although given the large uncertainty we treat this source as a non-detection. \\
\textit{HD282276}, \textit{HD61712}, and \textit{HD158643}. These sources show no detectable companions or polarized signal.

\subsection{Other programs}
The SPHERE image of eight sources from program \mbox{0101.C-0383} (PI: de Boer) has not been published before. The program was devoted to the characterization of disks around massive B stars. Four of these eight sources have been published by the GPI team \citep{Rich2022}, namely HD38087, HD53367, HD176386 (all stellar systems with undetected polarized light), and MWC297 (detected in scattered light from both the GPI and SPHERE images). Additionally, a prominent arc-like structure is detected around HD58647 with NaCo by \citet{Cugno2023}. The same structure is marginally visible from the SPHERE image as well. The remaining three sources of de Boer’s program are two non-detections (PDS 174, HD36982) and one marginal detection (EM\mbox{*} LkH$\alpha$ 215, see Fig.\,\ref{fig:new_sources_gallery}), where some signal is visible to South out to 0.3\arcsec. Given our distance to the source (739 pc), this feature extends out to 200 au from the star, and is realistically part of the disk since the millimeter flux is extremely high (58 mJy) for a source so far out.  

Three sources from the SPHERE GTO (1100.C-0481, \mbox{PI: Beuzit}) are also unpublished. TY CrA was also observed by GPI \citep{Rich2022} where it shows some very extended environmental structures that are also promptly detected with SPHERE (see Fig.\,\ref{fig:new_sources_gallery}). To our knowledge, no NIR image of EM\mbox{*} SR13 and EM\mbox{*} LkH$\alpha$348 has been published before. The image of EM\mbox{*} SR13 reveals a stellar companion at 0.24\arcsec\ (28 au) and a faint, compact emission polarimetric emission centered on the primary star. In addition, an extended feature in the visual direction of the companion is visible at projected separations larger than the companion itself (see Fig.\,\ref{fig:new_sources_gallery}). In LkH$\alpha$348, some polarized signal is detected to the SW of the star only. Given the very large distance to this source (more than 2 kpc), this signal would come from separations of several hundred au from the star. Similarly to the case of LkH$\alpha$215, concluding that this is the detection of a large disk is reasonable since the observed millimeter flux is very high (88 mJy).

Finally, the sample includes two sources, 2MASS J16064794-1841437 (shortened J1606-1841) and DoAr 28 (alias Haro 1-8), from program 0103.C-0470 (PI: Benisty) and two sources, Sz 111 and HD145718 (V718 Sco), from program 099.C-0891 (PI: Benisty). HD145718 was observed by GPI \citep{Davies2022} revealing a very inclined disk that looks very similar in the SPHERE image. DoAr 28 was observed with HiCiao by \citet{Rich2015} who reported polarized emission from an inclined disk and from some asymmetric features. The SPHERE image reveals the same inclined disk but shows no obvious evidence of extended emission exterior to it. The SPHERE image of J1606-1841 is, to our knowledge, the first NIR image of the source and it shows an extended, inclined disk with a P.A.\ of 10\degree\ (Fig.\,\ref{fig:new_sources_gallery}). A radial discontinuity is visible at 0.25\arcsec\ between an inner and an outer disk. Finally, the original SPHERE image of Sz 111 shows a relatively inclined disk with a P.A.\ of 45\degree. The disk brightness is decreased inward of 0.2\arcsec\ down to coronagraph indicating a possible disk cavity of approximately 30 au in radius.

\begin{table*}
\caption{Sample properties (targets 1-55). Columns are source name, high-contrast image type (see Sect.\,\ref{sec:analysis_scrutiny}), disk features {(C for cavity, S for spiral(s), R for ring(s), B for back side, BS for broad shadow(s), and NS for narrow shadow(s))}, ambient features {(E for envelope, B for binary interaction, A for ambient interaction with the environment)}, contrast $\alpha_{\rm pol}$, earliest reference for the image (or observing program if unpublished).}
\label{tab:sample_r1_g3}
\centering
\begin{tabular}{cccccc}
\toprule
            Source name &    Image type &       Disk Features & Ambient Features & $\alpha_{\rm pol}\ (\cdot10^{-3})$ &                   Reference \\
\midrule
                HD19745 & Non-detection &          \ &        \ &               0 &             0111.C-0369, Vioque \\
2MASS J03254982+3110237 &    Faint disk &          \ &        \ &             0.8 $\pm$ 0.2 &             0111.C-0369, Vioque \\
            V* V918 Per & Non-detection &          \ &        \ &               0 &             0111.C-0369, Vioque \\
            EM* LkHa330 &   Bright disk &          S &        \ &             4.5 $\pm$ 0.7 &         \citet{Pinilla2022} \\
                 CW Tau &    Faint disk &          \ &        \ &             1.5 $\pm$ 0.3 &          \citet{Garufi2024} \\
                 CX Tau &    Faint disk &          \ &        \ &             0.4 $\pm$ 0.1 &          \citet{Garufi2024} \\
                 CY Tau &    Faint disk &          \ &        \ &             0.6 $\pm$ 0.2 &             \citet{Ren2023} \\
               V892 Tau &    Faint disk &          \ &        \ &               \ &           \citet{Cugno2023} \\
               V409 Tau &    Faint disk &          B &        \ &             2.9 $\pm$ 0.6 &          \citet{Garufi2024} \\
                 BP Tau &    Faint disk &         BS &        \ &             0.5 $\pm$ 0.3 &          \citet{Garufi2024} \\
                 DE Tau &    Faint disk &          \ &        \ &             0.6 $\pm$ 0.2 &          \citet{Garufi2024} \\
                 RY Tau &       Ambient &          \ &     E. A &               \ &          \citet{Takami2013} \\
                  T Tau &       Ambient &          \ &     B. A &               \ &            \citet{Yang2018} \\
                 IP Tau &    Faint disk &          S &        \ &             1.3 $\pm$ 0.4 &             \citet{Ren2023} \\
                 DG Tau &       Ambient &      B. NS &     E. A &               \ &          \citet{Garufi2024} \\
                 DH Tau &    Faint disk &          \ &        \ &             0.9 $\pm$ 0.2 &    \citet{vanHolstein2021} \\
                 DI Tau & Non-detection &          \ &        \ &               0 &    \citet{deRegt2024} \\
                 IQ Tau &    Faint disk &          \ &        \ &             1.7 $\pm$ 0.3 &             \citet{Ren2023} \\
                 UX Tau &   Bright disk &          C &     B. A &             8.7 $\pm$ 0.7 &      \citet{Tanii2012} \\
                 DK Tau & Non-detection &          \ &        \ &               0 &          \citet{Garufi2024} \\
             ZZ Tau IRS &   Bright disk &          B &        \ &               \ & \citet{Hashimoto2024} \\
                 XZ Tau &       Ambient &          \ &     B. A &               \ &          \citet{Garufi2024} \\
               V710 Tau &    Faint disk &         BS &        \ &             2.7 $\pm$ 0.4 &          \citet{Garufi2024} \\
                 GG Tau &   Bright disk &   C. S. NS &     B. A &             8.7 $\pm$ 0.8 &       \citet{Itoh2014} \\
                 UZ Tau &    Faint disk &          \ &        \ &             2.5 $\pm$ 0.5 &          \citet{Garufi2024} \\
                HD28867 & Non-detection &          \ &        \ &               0 &             0111.C-0369, Vioque \\
               HD282276 & Non-detection &          \ &        \ &             0.1 $\pm$ 0.1 &             0111.C-0369, Vioque \\
               V807 Tau & Non-detection &          \ &        \ &               0 &          \citet{Garufi2024} \\
                 GI Tau &    Faint disk &          \ &        \ &             0.4 $\pm$ 0.2 &          \citet{Garufi2024} \\
                 GK Tau & Non-detection &          \ &        \ &               0 &          \citet{Garufi2024} \\
                 DL Tau &    Faint disk &          R &        \ &             0.7 $\pm$ 0.3 &             \citet{Ren2023} \\
                 HN Tau &    Faint disk &          \ &        \ &               \ &          \citet{Garufi2024} \\
                 DM Tau &    Faint disk &          C &        \ &             2.6 $\pm$ 0.4 &             \citet{Ren2023} \\
                 CI Tau &    Faint disk &          R &        \ &             1.4 $\pm$ 0.1 &     \citet{Garufi2024} 2021 \\
                 DN Tau &    Faint disk &          \ &        \ &             0.3 $\pm$ 0.2 &             \citet{Ren2023} \\
                 HQ Tau &    Faint disk &          \ &        \ &             0.4 $\pm$ 0.1 &          \citet{Garufi2024} \\
              V* HP Tau &       Ambient &          \ &        A &               \ &          \citet{Garufi2024} \\
           V* V1025 Tau & Non-detection &          \ &        \ &               0 &          \citet{Garufi2024} \\
                 DO Tau &       Ambient &          \ &     E. A &               \ &           \citet{Huang2022} \\
             EM* LkCa15 &   Bright disk &          C &        \ &             3.7 $\pm$ 0.5 &   \citet{Thalmann2016} \\
                 DQ Tau &    Faint disk &          \ &        \ &             0.4 $\pm$ 0.2 &          \citet{Garufi2024} \\
                 DR Tau &   Bright disk &          S &        A &               3.0 $\pm$ 0.5 &            \citet{Mesa2022} \\
                 DS Tau &    Faint disk &          \ &        \ &             0.2 $\pm$ 0.2 &             \citet{Ren2023} \\
                 UY Aur &       Ambient &          \ &     B. A &               \ &          \citet{Garufi2024} \\
                 GM Aur &   Bright disk &       S. B &        A &            16.4 $\pm$ 1.2 &        \citet{Oh2016}  \\
                HD31293 &   Bright disk &          S &        A &             9.7 $\pm$ 1.0 &  \citet{Hashimoto2011} \\
                 SU Aur &   Bright disk &      S. NS &        A &            10.1 $\pm$ 0.7 &    \citet{Jeffers2014} \\
                HD31648 &    Faint disk &          R &        \ &             0.5 $\pm$ 0.2 &   \citet{Kusakabe2012} \\
               V836 Tau &    Faint disk &          \ &        \ &             1.8 $\pm$ 0.5 &          \citet{Garufi2024} \\
                PDS 174 & Non-detection &          \ &        \ &               0 &            0101.C-0383, de Boer \\
                 RW Aur &       Ambient &          \ &     B. A &               \ &          \citet{Garufi2024} \\
              V1012 Ori &   Bright disk &          B &        \ &             3.9 $\pm$ 0.4 &        \citet{Valegard2024} \\
                 DW Aur &   Bright disk &          B &        \ &             4.7 $\pm$ 0.3 &               DESTINYS \\
                HD34282 &   Bright disk & C. S. R. B &        \ &             4.4 $\pm$ 0.5 &         \citet{deBoer2021} \\
                HD34700 &   Bright disk &   C. S. NS &        A &            12.7 $\pm$ 0.8 &   \citet{Monnier2019}       \\
\bottomrule
\end{tabular}
\end{table*}

\begin{table*}
\caption{Sample properties (targets 56-110). Continued from \ref{tab:sample_r1_g3}. Columns are source name, high-contrast image type (see Sect.\,\ref{sec:analysis_scrutiny}), disk features {(C for cavity, S for spiral(s), R for ring(s), B for back side, BS for broad shadow(s), and NS for narrow shadow(s))}, ambient features {(E for envelope, B for binary interaction, A for ambient interaction with the environment)}, contrast $\alpha_{\rm pol}$, earliest reference for the image (or observing program if unpublished).}
\label{tab:sample_r2_g3}
\centering
\begin{tabular}{cccccc}
\toprule
            Source name &    Image type &       Disk Features & Ambient Features & $\alpha_{\rm pol}\ (\cdot10^{-3})$ &                   Reference \\
\midrule
                             PDS 110 &    Faint disk &          \ &        \ &             0.3 $\pm$ 0.2  &        \citet{Valegard2024} \\
               HD287823 &    Faint disk &          \ &        \ &             0.5 $\pm$ 0.1 &   \citet{Garufi2022b}  \\
                PDS 111 &   Bright disk &       R. B &        \ &            14.1 $\pm$ 1.1 &         \citet{Derkink2024} \\
                PDS 113 &    Faint disk &          \ &        \ &             0.6 $\pm$ 0.3 &        \citet{Valegard2024} \\
                 GW Ori &   Bright disk &   C. S. BS &     B. A &               \ &           \citet{Kraus2020} \\
              V1650 Ori &    Faint disk &        \ &        \ &             0.2 $\pm$ 0.1 &     \citet{Valegard2024} \\
                HD36112 &   Bright disk &        S &        \ &             3.1 $\pm$ 0.4 & \citet{Benisty2015} \\
                 RY Ori & Non-detection &        \ &        \ &               0 &     \citet{Valegard2024} \\
2MASS J05341416-0536542 & Non-detection &        \ &        \ &               0 &     \citet{Valegard2024} \\
2MASS J05342495-0522055 & Non-detection &        \ &        \ &               0 &     \citet{Valegard2024} \\
            V* V543 Ori & Non-detection &        \ &        \ &               0 &     \citet{Valegard2024} \\
               HD245185 &    Faint disk &        R &        \ &               1.0 $\pm$ 0.3 &  \citet{Garufi2022b} \\
                HD36982 & Non-detection &        \ &        \ &               0 &         0101.C-0383, de Boer \\
              V1044 Ori & Non-detection &        \ &        \ &               0 &     \citet{Valegard2024} \\
                HD36917 & Non-detection &        \ &        \ &               0 &     \citet{Rich2022}     \\
              V2149 Ori & Non-detection &        \ &        \ &               0 &     \citet{Valegard2024} \\
                 NY Ori &    Faint disk &        \ &        \ &               \ &        \citet{Zurlo2024} \\
            V* V499 Ori & Non-detection &        \ &        \ &               0 &     \citet{Valegard2024} \\
                 CQ Tau &   Bright disk &    S. NS &        \ &             6.4 $\pm$ 0.6 &    \citet{Uyama2020}     \\
            V* V578 Ori & Non-detection &        \ &        \ &               0 &     \citet{Valegard2024} \\
               HD294260 &    Faint disk &        \ &        \ &             0.9 $\pm$ 0.5 &     \citet{Valegard2024} \\
               HD290770 & Non-detection &        \ &        \ &               0 &  \citet{Garufi2022b} \\
2MASS J05373003-0048520 & Non-detection &        \ &        \ &               0 &     \citet{Valegard2024} \\
              V1247 Ori &   Bright disk &        S &        \ &             4.8 $\pm$ 0.6 &   \citet{Ohta2016}  \\
           V* V1787 Ori & Non-detection &        \ &        \ &               0 &     \citet{Valegard2024} \\
               HD294268 &    Faint disk &        \ &        \ &             0.2 $\pm$ 0.2 &     \citet{Valegard2024} \\
                HD37411 & Non-detection &        \ &        \ &               0 &     \citet{Valegard2024} \\
                 TX Ori & Non-detection &        \ &        \ &               0 &     \citet{Valegard2024} \\
               V599 Ori &   Bright disk &     R. B &        \ &             3.5 $\pm$ 0.5 &     \citet{Valegard2024} \\
                 RV Ori &    Faint disk &        \ &        \ &               2.0 $\pm$ 0.5 &     \citet{Valegard2024} \\
               V606 Ori &    Faint disk &        \ &        \ &               1.0 $\pm$ 0.2 &     \citet{Valegard2024} \\
              Haro 5-38 & Non-detection &        \ &        \ &               0 &     \citet{Valegard2024} \\
                HD37806 &    Faint disk &        \ &        \ &               \ &     \citet{Rich2022}     \\
                HD38087 & Non-detection &        \ &        \ &               0 &     \citet{Rich2022}     \\
               V351 Ori &   Bright disk &        S &        \ &             5.6 $\pm$ 0.6 &    \citet{Wagner2020}   \\
                 FU Ori &       Ambient &        \ &     B. A &               \ &    \citet{Liu2016}  \\
              V1647 Ori & Non-detection &        \ &        \ &               0 & \citet{deRegt2024} \\
               HD250550 &       Ambient &        \ &     B. A &               \ &          \citet{Laws2020} \\
                HD45677 &       Ambient &        \ &     E. A &               \ &          \citet{Laws2020} \\
            EM* LkHa215 &    Faint disk &        \ &        \ &               \ &         0101.C-0383, de Boer \\
               HD259431 &       Ambient &        \ &     E. A &               \ &     \citet{Rich2022}     \\
                  R Mon &       Ambient &        \ &        A &               \ & \citet{deRegt2024} \\
                HD50138 &       Ambient &        \ &  E. B. A &               \ &          0111.C-0369, Vioque \\
               V960 Mon &       Ambient &        \ &     B. A &               \ &        \citet{Zurlo2024} \\
                  Z CMa &       Ambient &        \ &     B. A &               \ &    \citet{Liu2016}  \\
                HD53367 & Non-detection &        \ &        \ &               0 &    \citet{Rich2022}      \\
                 NX Pup & Non-detection &        \ &        \ &               0 & \citet{deRegt2024} \\
                HD58647 &   Bright disk &        \ &        \ &               \ &        \citet{Cugno2023} \\
                HD61712 & Non-detection &        \ &        \ &               0 &          0111.C-0369, Vioque \\
               V646 Pup & Non-detection &        \ &        \ &               0 & \citet{deRegt2024} \\
                PDS 277 &    Faint disk &        \ &        \ &             0.8 $\pm$ 0.2 &            DESTINYS \\
                 AT Pyx &   Bright disk & C. S. NS &        A &            25.2 $\pm$ 1.8 &       \citet{Ginski2022} \\
                HD72106 &   Bright disk &        \ &        \ &               \ &        \citet{Cugno2023} \\
                 ET Cha &    Faint disk &        \ &        \ &               \ &       \citet{Ginski2020} \\
                 EP Cha & Non-detection &        \ &        \ &               0 &            DESTINYS \\
              
\bottomrule
\end{tabular}
\end{table*}

\begin{table*}
\caption{Sample properties (targets 111-165). Continued from \ref{tab:sample_r1_g3}. Columns are source name, high-contrast image type (see Sect.\,\ref{sec:analysis_scrutiny}), disk features {(C for cavity, S for spiral(s), R for ring(s), B for back side, BS for broad shadow(s), and NS for narrow shadow(s))}, ambient features {(E for envelope, B for binary interaction, A for ambient interaction with the environment)}, contrast $\alpha_{\rm pol}$, earliest reference for the image (or observing program if unpublished).}
\label{tab:sample_r3_g3}
\centering
\begin{tabular}{cccccc}
\toprule
                 Source name &    Image type &       Disk Features & Ambient Features & $\alpha_{\rm pol}\ (\cdot10^{-3})$ &                   Reference \\
\midrule
                      Hen 3-225 & Non-detection &        \ &        \ &               0 &         \citet{Rich2022} \\
                HD85567 &    Faint disk &        \ &        \ &               \ &     \citet{Rich2022}     \\
                HD87643 &       Ambient &        \ &     B. A &               \ &          \citet{Laws2020} \\
            Wray 15-535 &    Faint disk &        \ &        \ &               \ &         \citet{Rich2022} \\
                 SY Cha &    Faint disk &        \ &        \ &             0.8 $\pm$ 0.1 &          \citet{Ren2023} \\
                 SZ Cha &   Bright disk &        R &        \ &             3.8 $\pm$ 0.3 &          \citet{Ren2023} \\
                 TW Cha &    Faint disk &        \ &        \ &             0.5 $\pm$ 0.2 &  \citet{Garufi2022b} \\
                 CR Cha &    Faint disk &        \ &        \ &             1.1 $\pm$ 0.3 &       \citet{Ginski2024} \\
                 TW Hya &   Bright disk &        R &        \ &             7.2 $\pm$ 0.6 &  \citet{Rapson2015} \\
                HD95881 &    Faint disk &        \ &        \ &               \ &     \citet{Rich2022}     \\
                      CS Cha &   Bright disk &           \ &        \ &             4.4 $\pm$ 0.3 &         \citet{Ginski2024} \\
                      CT Cha &    Faint disk &           \ &        \ &             0.4 $\pm$ 0.1 &         \citet{Ginski2024} \\
             RX J1106.3-7721 & Non-detection &           \ &        \ &               0 &         \citet{Ginski2024} \\
                      DI Cha & Non-detection &           \ &        \ &               0 &    \citet{Garufi2022b} \\
                     HD97048 &   Bright disk &           R &        \ &             3.4 $\pm$ 0.4 &    \citet{Ginski2016} \\
                      HP Cha &    Faint disk &           \ &     B. A &             1.1 $\pm$ 0.2 &          \citet{Zhang2023b} \\
                      VZ Cha &    Faint disk &           \ &        \ &             0.8 $\pm$ 0.1 &         \citet{Ginski2024} \\
                      WX Cha & Non-detection &           \ &        \ &               0 &         \citet{Ginski2024} \\
                      WW Cha &   Bright disk &           S &        A &             3.9 $\pm$ 0.4 &         \citet{Garufi2020a} \\
                      WY Cha & Non-detection &           \ &        \ &               0 &         \citet{Ginski2024} \\
                      TWA 3A & Non-detection &           \ &        \ &               0 &   \citet{deRegt2024} \\
                      PDS 51 & Non-detection &           \ &        \ &               0 &         \citet{Ginski2024} \\
                     CHX 18N & Non-detection &           \ &        \ &               0 &         \citet{Ginski2024} \\
                       Sz 41 & Non-detection &           \ &        \ &               0 &         \citet{Ginski2024} \\
                      CV Cha &    Faint disk &           \ &        \ &             1.1 $\pm$ 0.2 &         \citet{Ginski2024} \\
                      CHX 22 &       Ambient &           \ &     B. A &               \ &          \citet{Zhang2023b} \\
                       Sz 45 &    Faint disk &           \ &        \ &             0.7 $\pm$ 0.2 &         \citet{Garufi2020a} \\
                 Wray 15-788 &   Bright disk &       R. BS &        \ &             9.1 $\pm$ 0.6 &      \citet{Bohn2019} \\
                     HD98800 &    Faint disk &           \ &        \ &               \ &       \citet{Rich2022}     \\
                     HD98922 &    Faint disk &           \ &        \ &             0.2 $\pm$ 0.2 &    \citet{Garufi2022b} \\
                    HD100453 &   Bright disk &       S. NS &     B. A &             5.3 $\pm$ 0.3 &    \citet{Wagner2015} \\
                    HD100546 &   Bright disk &        C. S &        A &             3.6 $\pm$ 0.5 &    \citet{Garufi2016} \\
                    HD101412 & Non-detection &           \ &        \ &               0 &           \citet{Rich2022} \\
                      DZ Cha &   Bright disk &           S &        \ &             4.1 $\pm$ 0.4 &   \citet{Canovas2018} \\
                       T Cha &   Bright disk &           \ &        \ &            13.7 $\pm$ 0.7 &   \citet{Pohl2017} \\
                    HD104237 &    Faint disk &           \ &        \ &               \ &           \citet{Rich2022} \\
Gaia DR2 5854897321965963264 &    Faint disk &           R &        \ &             2.2 $\pm$ 0.4 &              DESTINYS \\
                      AX Cha &       Ambient &           \ &     E. A &               \ &              DESTINYS \\
             IRAS 12535-7623 & Non-detection &           \ &        \ &               0 &              DESTINYS \\
     2MASS J13005622-7654021 & Non-detection &           \ &        \ &               0 &              DESTINYS \\
                       Sz 51 &    Faint disk &           \ &        \ &             0.4 $\pm$ 0.3 &              DESTINYS \\
                      CM Cha &    Faint disk &           \ &        \ &             1.2 $\pm$ 0.4 &              DESTINYS \\
                       Hn 23 & Non-detection &           \ &        \ &               0 &              DESTINYS \\
                       Hn 24 & Non-detection &           \ &        \ &               0 &              DESTINYS \\
                   Hen 3-854 & Non-detection &           \ &        \ &               0 &              DESTINYS \\
                      PDS 66 &   Bright disk &           R &        \ &             4.6 $\pm$ 0.3 &     \citet{Wolff2016} \\
     2MASS J13342461-6517473 &    Faint disk &           R &        \ &             2.3 $\pm$ 0.2 &              DESTINYS \\
                      PDS 70 &   Bright disk &           C &        \ &             9.7 $\pm$ 0.6 & \citet{Hashimoto2011} \\
                   HD135344B &   Bright disk &    C. S. NS &        \ &             9.4 $\pm$ 0.5 &   \citet{Muto2012} \\
                      IK Lup &    Faint disk &           B &        \ &             0.7 $\pm$ 0.3 &         \citet{Garufi2020a} \\
                    HD139614 &   Bright disk &    C. R. BS &        \ &             4.3 $\pm$ 0.5 &     \citet{MuroArena2020} \\
                      HT Lup &    Faint disk &           \ &        \ &             1.7 $\pm$ 0.3 &         \citet{Garufi2020a} \\
                      GW Lup &    Faint disk &           \ &        \ &             0.5 $\pm$ 0.1 &    \citet{Garufi2022b} \\
                       Sz 72 &    Faint disk &           \ &        \ &               \ &              DESTINYS \\
                      GQ Lup &   Bright disk &           S &     B. A &             3.8 $\pm$ 0.5 &   \citet{vanHolstein2021} \\
\bottomrule
\end{tabular}
\end{table*}

\begin{table*}
\caption{Sample properties (targets 166-220). Continued from \ref{tab:sample_r1_g3}. Columns are source name, high-contrast image type (see Sect.\,\ref{sec:analysis_scrutiny}), disk features {(C for cavity, S for spiral(s), R for ring(s), B for back side, BS for broad shadow(s), and NS for narrow shadow(s))}, ambient features {(E for envelope, B for binary interaction, A for ambient interaction with the environment)}, contrast $\alpha_{\rm pol}$, earliest reference for the image (or observing program if unpublished).}
\label{tab:sample_r4_g3}
\centering
\begin{tabular}{cccccc}
\toprule
            Source name &    Image type &       Disk Features & Ambient Features & $\alpha_{\rm pol}\ (\cdot10^{-3})$ &                   Reference \\
\midrule
PDS 144 A &    Faint disk &          NS &        \ &             2.5 $\pm$ 0.3 &              DESTINYS \\
                    HD141569 &    Faint disk &        C. R &        \ &               \ &    \citet{Perrot2016} \\
                    HD141637 & Non-detection &           \ &        \ &               0 &              DESTINYS \\
                       Sz 77 & Non-detection &           \ &        \ &               0 &              DESTINYS \\
                WRAY 15-1384 &   Bright disk &           B &        \ &             3.3 $\pm$ 0.5 &              DESTINYS \\
                      IM Lup &   Bright disk &        R. B &        \ &             9.3 $\pm$ 0.7 &       \citet{Avenhaus2018} \\
                    HD142666 &    Faint disk &           \ &        \ &             1.4 $\pm$ 0.2 &    \citet{Garufi2022b} \\
                    HD142527 &   Bright disk & C. S. B. NS &        A &            20.6 $\pm$ 1.1 &  \citet{Canovas2013} \\
                      RU Lup &    Faint disk &           \ &        \ &             1.1 $\pm$ 0.2 &       \citet{Avenhaus2018} \\
                      Sz 126 & Non-detection &           \ &        \ &               0 &              DESTINYS \\
     2MASS J15575444-2450424 &    Faint disk &           \ &        \ &             2.2 $\pm$ 0.2 &              DESTINYS \\
                    HD143006 &   Bright disk &    C. R. BS &        \ &             3.5 $\pm$ 0.3 &   \citet{Benisty2018} \\
                      RY Lup &   Bright disk &           S &        \ &            17.5 $\pm$ 0.7 &       \citet{Langlois2018} \\
                      MY Lup &   Bright disk &           B &        \ &            14.7 $\pm$ 0.7 &       \citet{Avenhaus2018} \\
                      EX Lup &    Faint disk &          BS &        A &               \ &      \citet{Rigliaco2020}  \\
2MASS J16035767-2031055 & Non-detection &        \ &        \ &               0 &        \citet{Garufi2020a} \\
        RX J1604.3-2130 &   Bright disk &    C. NS &        \ &            18.8 $\pm$ 1.6 &  \citet{Pinilla2015b} \\
2MASS J16062196-1928445 & Non-detection &        \ &        \ &               0 &        \citet{Garufi2020a} \\
2MASS J16064385-1908056 & Non-detection &        \ &        \ &               0 &        \citet{Garufi2020a} \\
2MASS J16064794-1841437 &   Bright disk &        B &        \ &               9.0 $\pm$ 0.7 &          0103.C-0470, Benisty \\
               HD144432 &    Faint disk &       BS &        \ &             0.5 $\pm$ 0.2 &   \citet{Garufi2022b} \\
            V* HO Lup A &    Faint disk &        \ &        \ &             0.3 $\pm$ 0.1 &             DESTINYS \\
                  Sz 91 &    Faint disk &        \ &        \ &               \ &    \citet{Tsukagoshi2014} \\
                 HK Lup & Non-detection &        \ &        \ &               0 &        \citet{Garufi2020a} \\
2MASS J16082324-1930009 &   Bright disk &        B &        \ &             3.2 $\pm$ 0.4 &             DESTINYS \\
2MASS J16083070-3828268 &   Bright disk &        B &        \ &            18.4 $\pm$ 0.9 &     \citet{Villenave2019} \\
               HD144668 & Non-detection &        \ &        \ &               0 &   \citet{Garufi2017} \\
              V1094 Sco &   Bright disk &        R &        \ &            10.2 $\pm$ 0.6 &        \citet{Garufi2020a} \\
                 Sz 111 &   Bright disk &        C &        \ &               6.0 $\pm$ 0.6 &          099.C-089, Benisty \\
2MASS J16090075-1908526 &    Faint disk &        R &        \ &             2.1 $\pm$ 0.4 &        \citet{Garufi2020a} \\
                 Sz 134 &    Faint disk &        \ &        \ &               \ &             DESTINYS \\
        IRAS 16072-2057 &   Bright disk &     C. B &        \ &            20.5 $\pm$ 1.1 &             DESTINYS \\
2MASS J16102857-1904469 & Non-detection &        \ &        \ &               0 &        \citet{Garufi2020a} \\
               HD145263 & Non-detection &        \ &        \ &               0 &   \citet{Garufi2017} \\
2MASS J16111534-1757214 & Non-detection &        \ &        \ &               0 &        \citet{Garufi2020a} \\
              EM* AS205 &    Faint disk &        \ &     E. A &             1.8 $\pm$ 0.7 &         \citet{Weber2023} \\
2MASS J16120668-3010270 &   Bright disk &     C. S &        \ &             6.9 $\pm$ 0.7 &        \citet{Ginski2024} \\
               HD145718 &    Faint disk &        B &        \ &             2.2 $\pm$ 0.3 &        \citet{Davies2022} \\
              V1098 Sco &   Bright disk & S. R. BS &        \ &            16.4 $\pm$ 0.6 &             \citet{Williams2025} \\
2MASS J16141107-2305362 & Non-detection &        \ &        \ &               0 &        \citet{Garufi2020a} \\
2MASS J16142029-1906481 &    Faint disk &       BS &        \ &             1.1 $\pm$ 0.2 &        \citet{Garufi2020a} \\
        RX J1615.3-3255 &   Bright disk &     R. B &        \ &            13.3 $\pm$ 0.8 &  \citet{deBoer2016} \\
                 VV Sco & Non-detection &        \ &        \ &               0 &        \citet{Garufi2020a} \\
2MASS J16154416-1921171 &   Bright disk &        \ &        A &             5.6 $\pm$ 0.5 &        \citet{Garufi2020a} \\
                PDS 415 &   Bright disk &        B &        \ &            15.8 $\pm$ 0.5 &             DESTINYS \\
               Haro 1-1 &   Bright disk &     R. B &        \ &             5.2 $\pm$ 0.4 &             DESTINYS \\
               V935 Sco &   Bright disk &        B &        A &             4.9 $\pm$ 0.6 &             DESTINYS \\
                DoAr 16 & Non-detection &        \ &        \ &               0 &        \citet{Garufi2020a} \\
                EM* SR4 &    Faint disk &        R &        \ &             1.1 $\pm$ 0.3 &        \citet{Garufi2020a} \\
                DoAr 21 &       Ambient &        \ &        A &               \ &        \citet{Garufi2020a} \\
                DoAr 24 &    Faint disk &        \ &        \ &               \ &  \citet{deRegt2024} \\
                DoAr 25 &   Bright disk &        B &        \ &             8.4 $\pm$ 0.4 &        \citet{Garufi2020a} \\
                 ROX 12 & Non-detection &        \ &        \ &               0 &  \citet{deRegt2024} \\
                DoAr 28 &    Faint disk &        \ &        \ &             1.6 $\pm$ 0.3 &  \citet{Rich2015}    \\
2MASS J16265280-2343127 & Non-detection &        \ &        \ &               0 &             DESTINYS \\
               
\bottomrule
\end{tabular}
\end{table*}

\begin{table*}
\caption{Sample properties (targets 221-268). Continued from \ref{tab:sample_r1_g3}. Columns are source name, high-contrast image type (see Sect.\,\ref{sec:analysis_scrutiny}), disk features {(C for cavity, S for spiral(s), R for ring(s), B for back side, BS for broad shadow(s), and NS for narrow shadow(s))}, ambient features {(E for envelope, B for binary interaction, A for ambient interaction with the environment)}, contrast $\alpha_{\rm pol}$, earliest reference for the image (or observing program if unpublished).}
\label{tab:sample_r5_g3}
\centering
\begin{tabular}{cccccc}
\toprule
            Source name &    Image type &       Disk Features & Ambient Features & $\alpha_{\rm pol}\ (\cdot10^{-3})$ &                   Reference \\
\midrule
            EM* SR23 & Non-detection &        \ &        \ &               0 &             DESTINYS \\
             EM* SR24 S &       Ambient &        \ &  E. B. A &               \ &       \citet{Mayama2020}  \\
               EM* SR21 &   Bright disk &     S. R &        \ &             3.9 $\pm$ 0.5 & \citet{Follette2013} \\
               EM* SR12 & Non-detection &        \ &        \ &               0 &  \citet{deRegt2024} \\
               WLY 2-48 &   Bright disk &        C &        \ &               \ & \citet{Follette2015} \\
                EM* SR9 & Non-detection &        \ &        \ &               0 &        \citet{Garufi2020a} \\
               EM* SR13 &    Faint disk &        \ &     B. A &             1.9 $\pm$ 0.3 &                  SPHERE GTO \\
               EM* SR20 & Non-detection &        \ &        \ &               0 &           \citet{Ren2023} \\
               ROXs 42C & Non-detection &        \ &        \ &               0 &  \citet{deRegt2024} \\
                DoAr 44 &   Bright disk &    C. NS &        A &             8.7 $\pm$ 0.7 &      \citet{Avenhaus2018} \\
           V* V1003 Oph &    Faint disk &       BS &        \ &             0.2 $\pm$ 0.1 &   \citet{Garufi2022b} \\
                 WSB 82 &    Faint disk &       BS &        \ &             0.4 $\pm$ 0.1 &   \citet{Garufi2022b} \\
               HD150193 & Non-detection &        \ &        \ &               0 &   \citet{Garufi2022b} \\
               Wa Oph 6 &    Faint disk &        S &        \ &             0.4 $\pm$ 0.3 & \citet{Brown-Sevilla2021} \\
              EM* AS209 &    Faint disk &        R &        \ &             2.2 $\pm$ 0.5 &      \citet{Avenhaus2018} \\
               HD152404 &   Bright disk &       NS &        \ &             4.3 $\pm$ 0.4 &   \citet{Garufi2022b} \\
               V921 Sco &    Faint disk &        \ &        \ &               \ &          \citet{Rich2022} \\
             Hen 3-1330 & Non-detection &        \ &        \ &               0 &          \citet{Rich2022} \\
              EM* AS218 &    Faint disk &        R &        \ &             1.8 $\pm$ 0.3 &             DESTINYS \\
              V* KK Oph &   Bright disk &        B &        \ &             4.8 $\pm$ 0.6 &             DESTINYS \\
            EM* LkHa345 &    Faint disk &     \ &        \ &             1.5 $\pm$ 0.4 &             DESTINYS \\
2MASS J17110392-2722551 &   Bright disk & C. BS &        A &               \ &         \citet{Zurlo2024} \\
                PDS 453 &   Bright disk &     B &        \ &            24.2 $\pm$ 1.5 &     \citet{Martinien2024} \\
2MASS J17285631-2710031 & Non-detection &     \ &        \ &               0 &             DESTINYS \\
               HD319896 & Non-detection &     \ &        \ &               0 &  \citet{deRegt2024} \\
               HD158643 & Non-detection &     \ &        \ &               0 &          \citet{Rich2022} \\
               HD163296 &    Faint disk &     R &        \ &             0.9 $\pm$ 0.2 &   \citet{Garufi2014b} \\
              V4046 Sgr &   Bright disk & R. NS &        \ &             8.2 $\pm$ 0.6 &  \citet{Rapson2015} \\
               HD169142 &   Bright disk &  C. R &        \ &             4.2 $\pm$ 0.4 &  \citet{Quanz2013b}  \\
            EM* MWC 297 &   Bright disk &     \ &        \ &               \ &     \citet{Rich2022}      \\
                 VV Ser & Non-detection &     \ &        \ &               \ &           0111.C-0369, Vioque \\
            EM* LkHa348 &    Faint disk &    BS &        \ &             0.5 $\pm$ 0.1 &                  SPHERE GTO \\
                 BN CrA &   Bright disk &  R. B &        \ &               4.0 $\pm$ 0.3 &   \citet{Columba2025} \\
        RX J1842.9-3532 &   Bright disk &     S &        \ &             3.5 $\pm$ 0.4 &             DESTINYS \\
        RX J1852.3-3700 &    Faint disk &     R &        \ &             2.3 $\pm$ 0.3 &     \citet{Villenave2019} \\
2MASS J19005804-3645048 &    Faint disk &     B &        \ &             1.8 $\pm$ 0.3 &             DESTINYS \\
                  S CrA &       Ambient &     \ &     B. A &               \ &         \citet{Zhang2023b} \\
               HD176386 & Non-detection &     \ &        \ &               0 &      \citet{Rich2022}     \\
                 TY CrA &       Ambient &     \ &     B. A &               \ &     \citet{Rich2022}      \\
                  R CrA &       Ambient &     \ &     B. A &               \ & \citet{Mesa2019} \\
                 DG CrA &   Bright disk &    NS &     B. A &            18.2 $\pm$ 0.7 &             DESTINYS \\
                  T CrA &   Bright disk &     \ &        A &            26.6 $\pm$ 1.0 &      \citet{Rigliaco2023} \\
                 VV CrA &       Ambient &     \ &     B. A &               \ &             DESTINYS \\
            V* V721 CrA &   Bright disk &     B &        A &             7.1 $\pm$ 0.9 &   \citet{Columba2025} \\
               HD179218 &    Faint disk &     \ &        \ &             2.3 $\pm$ 0.3 &   \citet{Garufi2022b} \\
              V1295 Aql & Non-detection &     \ &        \ &               0 &          \citet{Rich2022} \\
              V1057 Cyg &       Ambient &     \ &        A &               \ &      \citet{Liu2016} \\
              V1735 Cyg &       Ambient &     \ &        A &               \ &      \citet{Liu2016} \\
\bottomrule
\end{tabular}
\end{table*}

\end{document}